\documentclass[11pt,a4paper]{article}
\usepackage{amssymb}
\usepackage{amsfonts}
\usepackage{amsmath}
\usepackage{amsfonts}
\usepackage{amsmath,amssymb}
\usepackage{epsfig,graphicx}

\input{}
\begin{document}

\begin{center}
{\Large Gravity and Unification: Insights from SL(2N,C) Gauge Theories}

{\huge \ }

{\Huge \bigskip }

\bigskip

\bigskip

\textbf{J. L.~Chkareuli}$^{1,2}$

\bigskip

$^{1}$\textit{Institute of Theoretical\ Physics, Ilia State University, 0179
Tbilisi, Georgia}

$^{2}$\textit{Andronikashvili} \textit{Institute of Physics, Tbilisi State
University, 0177 Tbilisi, Georgia\ \ }

\bigskip {\scriptsize \bigskip }\bigskip \bigskip \bigskip \bigskip \bigskip

\textbf{Abstract}
\end{center}

\bigskip

The perspective that gravity may govern the unification of all elementary
forces calls for extending the gauge-gravity symmetry $SL(2,C)$ to the
broader local symmetry $SL(2N,C)$, where $N$ reflects the internal $SU(N)$
subgroup. This extension yields a consistent hyperunification framework in
which---aside from the linear gravity Lagrangian, to which only tensor
fields contribute---the quadratic curvature sector is fully unified across
all gauge submultiplets. Tetrad fields play a central role: once dynamical,
their invertibility---treated as a nonlinear sigma-model type length
constraint---naturally implies condensation and thereby triggers spontaneous
breaking of $SL(2N,C)$. As a result, while the full gauge multiplet contains
vector, axial-vector, and tensor submultiplets, only the vector submultiplet
remains in the observed spectrum; the axial-vector and tensor submultiplets
acquire large masses at the symmetry-breaking scale. The effective symmetry
reduces to $SL(2,C)\times SU(N)$, collecting together $SL(2,C)$ gauge
gravity and the $SU(N)$ grand-unified sector. Since states in $SL(2N,C)$ are
also classified by spin, many $SU(N)$ GUT models---such as standard $SU(5)$%
---appear ill-suited for fundamental spin-$1/2$ quarks and leptons. By
contrast, applying $SL(2N,C)$ to a composite framework with chiral preons in
fundamental representations points to $SL(16,C)$, with effective $%
SL(2,C)\times SU(8)$ accommodating all three quark-lepton families, as a
compelling candidate for hyperunification.

%
\thispagestyle{empty}\newpage

\section{Introduction}

A deepening recognition has long emerged that gravity exhibits notable
similarities to the other three fundamental forces when examined within the
framework of conventional gauge theory \cite{Utiyama, Kibble, ish}.
Specifically, the spin-connection fields that gauge Lorentz symmetry arise
in a manner analogous to photons and gluons in the Standard Model. This
raises the intriguing possibility that these spin connections could be
unified with the Standard Model gauge bosons within a larger non-compact
symmetry group, potentially leading to a unified description of all known
gauge interactions.

In discussing Lorentz gauge gravity, we focus on the $SL(2,C)$ symmetry
group (and its extensions) rather than the conventional $SO(1,3)$, since $%
SL(2,C)$ more fundamentally captures the spinorial nature of fermions in
spacetime. Accordingly, if we accept that gravity itself governs the
unification of all elementary forces, then the natural extension of the
gauge gravity group $SL(2,C)$ leads to the broader local symmetry $SL(2N,C)$%
, where $N$ defines the degree of the internal $SU(N)$ symmetry subgroup. In
this context, we refer to such frameworks as hyperunified theories (HUTs),
with particular emphasis on the $SL(2N,C)$ model examined here in detail.
This unification integrates $SL(2,C)$ gauge gravity with $SU(N)$ grand
unified theory. The $SU(N)$ subgroup within $SL(2N,C)$ is designated as a
"hyperflavor" symmetry, in contrast to the \textquotedblleft
neutral\textquotedblright\ $SL(2,C)$ of pure gravity. Remarkably, the total
gauge multiplet of the $SL(2N,C)$ HUT may encompass both spin-$1$ and spin-$2
$ fields---organized into vector, axial-vector, and tensor submultiplets---
offering a potential unified framework for the Standard Model and gravity,
provided several familiar challenges can be addressed.

Numerous models in the literature propose unifying gravity with other
interactions by merging local Lorentz and internal symmetries into a
non-compact covering group \cite{ish1, cho, hu, per, cham, N, jpl}. Among
these, the $SL(2N,C)$ HUTs---despite their appeal---have been comparatively
less explored, owing to certain generic difficulties.

First, the vector fields in the full gauge multiplet are invariably
accompanied by axial-vector and tensor partners, for which there is no
direct empirical evidence; most must be filtered out by direct constraints
or else decouple at symmetry breaking. Furthermore, while vectors are
intended to mediate ordinary gauge interactions, the tensor fields from the
same multiplet must furnish the subtle gravitational interactions consistent
with observation. Crucially, whereas in pure gravity one may work with an
action linear in the curvature ($R$) built from a tensor field, unification
with other interactions necessitates the inclusion of quadratic curvature ($%
R^{2}$) terms as well. Consequently, tensor fields in these terms will
induce interactions comparable to those of the gauge vector fields in the
Standard Model. Moreover, the tensor fields, akin to the vector ones,
exhibit now the internal $SU(N)$ symmetry features implying the existence of
the multiplet of hyperflavored gravitons rather than a single "neutral" one.
Apart from that, such $R+R^{2}$ Lagrangians for gravity are generally known
to contain ghosts and tachyons rendering them essentially unstable.

Second, the matter multiplets of $SL(2N,C)$---even at relatively low
dimension---are excessively large, containing hundreds of states of various
spins, and cannot, without additional constraints, be matched to the
observed quarks and leptons. Notably, $SL(6,C)$ symmetry \cite{ish1}---a
familiar case of $SL(2N,C)$---was explored in hadron physics in the late
1970s , with limited success. Applications to the Standard Model likewise
appear unpromising unless special constraints reveal a deeper level of
elementarity at which the framework can operate. This naturally points to a
composite picture in which quarks and leptons arise from chiral preons in
fundamental spin-$1/2$ representations, for which $SL(2N,C)$ provides a more
natural unifying framework. \

And finally, perhaps most importantly, hyperunified theories face the
Coleman--Mandula theorem \cite{18}, which forbids a nontrivial merging of
spacetime and internal symmetries within a conventional Lie algebra
framework. As is known, this theorem originally emerged in the context of
attempts to "relativize" the $SU(6)$ symmetry -- used for the spin-unitary
spin classification of mesons and baryons \cite{feza}\ -- by embedding it
into $SL(6,C)$.

By contrast, we show here that $SL(2N,C)$ may provide a natural resolution
of these challenges in a manner not previously explored. The key idea is
that the interactions of the extended gauge multiplet $I_{\mu }$ in the $%
SL(2N,C)$ theory is intrinsically shaped by the associated tetrads. Treating
the tetrads as dynamical fields, the usual invertibility (orthonormality)
condition---viewed as a nonlinear sigma--model type length
constraint---naturally implies their condensation, thereby triggering the
the $SL(2N,C)$ spontaneous violation. Consequently, while the full gauge
multiplet of $SL(2N,C)$ encompasses vector, axial-vector and tensor field
submultiplets, only the vector field submultiplet emerges in the observed
particle spectrum. The axial-vector and tensor field submultiplets acquire
heavy masses at the $SL(2N,C)$ symmetry breaking scale. Once the non-compact
internal directions decouple, the theory effectively exhibits a local $%
SL(2,C)\times SU(N)$ symmetry rather than the full $SL(2N,C)$ symmetry,
which solely serves to determine the structure of the gauge and matter
multiplets. Thus, tetrads not only fix the spacetime geometry but also
select which internal symmetries and associated gauge interactions are
operative in it.\ In this way, the restrictions imposed by the
Coleman-Mandula theorem appear naturally circumvented as well. \

Next, as mentioned, the extension of $SL(2,C)$ gauge gravity to $SL(2N,C)$
naturally calls for the inclusion of some "safe" quadratic curvature
(strength) terms in the gravitational sector of the extended theory,
alongside the standard quadratic strength-tensor terms for the vector
fields. We argue that this requirement uniquely identifies, from among all
possible candidates, the ghost-free curvature-squared gravity Lagrangian
initially proposed by Neville \cite{nev} (see also \cite{nev1}), as the most
appropriate model for such an extension. Consequently, the resulting theory
includes the $SL(2,C)$ symmetric $R+R%
{{}^2}%
$ Einstein-Cartan type gravity action, which remains free from ghosts and
tachyons, alongside the conventional $SU(N)$ gauge vector field theory,
considered a strong GUT candidate.

In this context, since states in $SL(2N,C)$ are also classified by their
spin, many $SU(N)$ grand-unified models---including standard $SU(5)$ \cite%
{gg} and its straightforward extensions---appear ill-suited for fundamental
spin-$1/2$ quarks and leptons. However, as argued above, in a composite
picture with constituent chiral preons the situation changes drastically. In
particular, the $SU(8)$ grand unification, arising from the $SL(16,C)$
symmetry---originally introduced for preons--- successfully describes all
three families of composite quarks and leptons. This follows as a
consequence of the conservation of the preon chiral symmetry $%
SU(8)_{L}\times SU(8)_{R}$ at large distances where the composites emerge.
In fact, the anomaly matching condition associated with this symmetry
conservation acts as the constraint that filters out all other states except
the physical quarks and leptons from the multiplets in which the composites
reside. This eventually results in a theory with the residual symmetry $%
SL(2,C)\times SU(8)$ for composite fermions, while the full $SL(16,C)$
symmetry remains intact for preons at small scales.

The paper is organized as follows. Section 2 revisits the $SL(2,C)$ gauge
gravity in an updated form. Section 3 presents the $SL(2N,C)$ HUT with its
breaking to the effective $SL(2,C)\times SU(N)$ symmetry triggered by the
tetrad condensation given in Section 4. Section 5 describes the hyperunified
linear and quadratic strength Lagrangians for gravity and other fundamental
forces, while Sections 6 and 7 highlight specific HUT models with a
particular focus on the $SL(16,C)$ theory with composite quarks and leptons.
The paper concludes with a summary in Section 8.

\section{$SL(2,C)$ gauge gravity}

We begin by presenting the $SL(2,C)$ gravity model, drawing in part from the
pioneering work \cite{ish}. One can assume that at each point in spacetime a
local frame exists, where the global $SL(2,C)$ symmetry group acts. Under
this symmetry, the fundamental fermions transform as%
\begin{equation}
\Psi \rightarrow \Omega \Psi \text{, \ \ }\Omega =\exp \left\{ \frac{i}{4}%
\theta _{ab}\gamma ^{ab}\right\} \text{ }  \label{om}
\end{equation}%
\newline
where the matrix $\Omega $ satisfies a pseudounitarity condition, $\Omega
^{-1}=\gamma _{0}\Omega ^{+}\gamma _{0}$, with the transformation parameters
$\theta _{ab}$ taken to be constant for now\footnote{%
The $SL(2,C)$ symmetry group, being the double cover of the proper
orthochronous Lorentz group $SO^{+}(1,3)$, can be naturally viewed as the
complexification of the familiar $SU(2)$ symmetry acting on chiral Weyl
spinors. Specifically, the $SL(2,C)$ algebra is spanned by the Pauli
matrices $\sigma _{i}$ and their complexified versions $i\sigma _{i}$
describing rotations and boosts, respectively. Meanwhile, the emergence of $%
SL(2,C)$ symmetry for Dirac spinors follows from combining left- and
right-handed Weyl spinors into a four-component fermion field $\Psi $ whose
block-diagonal transformation law (\ref{om}) ensures Lorentz invariance. In
a similar vein, the $SL(2N,C)$ symmetry group can be regarded as the
complexified counterpart of the unitary symmetry $SU(2N)$, a point that will
be relevant in later discussions.}. To maintain the invariance of the
kinetic terms, $i\overline{\Psi }\gamma ^{\mu }\partial _{\mu }\Psi $, the
gamma matrices must be replaced by a set of tetrad matrices $e^{\mu \text{ }}
$which transform like%
\begin{equation}
e^{\mu }\rightarrow \Omega e^{\mu }\Omega ^{-1}\text{ }  \label{trl}
\end{equation}%
In general, the tetrad matrices $e^{\mu \text{ }}$and their conjugates $%
e_{\mu }$ incorporate the appropriate tetrad fields $e_{a}^{\mu }$ and $%
e_{\mu }^{a}$, respectively,%
\begin{equation}
\text{\ }e^{\mu }=e_{a}^{\mu }\gamma ^{a}\text{ , \ }e_{\mu }=e_{\mu
}^{a}\gamma _{a}\text{ }  \label{tlf}
\end{equation}%
which transforms infinitesimally as
\begin{equation}
\delta e^{\mu c}=\frac{1}{2}\theta _{ab}(e^{\mu a}\eta ^{bc}-e^{\mu b}\eta
^{ac})  \label{tr1a}
\end{equation}%
They, as usual, satisfy the invertibility (or orthonormality) conditions%
\begin{equation}
e_{\mu }^{a}e_{a}^{\nu }=\delta _{\mu }^{\nu }\text{, \ }e_{\mu
}^{a}e_{b}^{\mu }=\delta _{b}^{a}  \label{or}
\end{equation}%
and determine the metric tensors in the theory
\begin{equation}
g_{\mu \nu }=\frac{1}{4}\mathrm{Tr}(e_{\mu }e_{\nu })=e_{\mu }^{a}e_{\nu
}^{b}\eta _{ab}\text{\ , \ }g^{\mu \nu }=\frac{1}{4}\mathrm{Tr}(e^{\mu
}e^{\nu })=e_{a}^{\mu }e_{b}^{\nu }\eta ^{ab}  \label{gmn}
\end{equation}

Turning now to the case where the $SL(2,C)$ transformations (\ref{om})
become local, $\theta _{ab}\equiv \theta _{ab}(x)$, one must introduce the
spin-connection gauge field multiplet $I_{\mu }$\ which transforms as usual%
\begin{equation}
I_{\mu }\rightarrow \Omega I_{\mu }\Omega ^{-1}-\frac{1}{ig}(\partial _{\mu
}\Omega )\Omega ^{-1}  \label{123}
\end{equation}%
This defines the covariant derivative for the fermion field
\begin{equation}
\partial _{\mu }\Psi \rightarrow D_{\mu }\Psi =\partial _{\mu }\Psi
+igI_{\mu }\Psi \text{ }  \label{124}
\end{equation}%
where $g$ is the gauge coupling constant. The multiplet $I_{\mu }$ gauging
the $SL(2,C)$ takes the following form when decomposed in global spacetime
\begin{equation}
I_{\mu }=\frac{1}{4}T_{\mu \lbrack ab]}\gamma ^{ab}\text{ \ \ }  \label{125}
\end{equation}%
where the gauge field components $T_{\mu \lbrack ab]}$ transform as%
\begin{equation}
\delta T_{\mu }^{[ab]}=\frac{1}{2}\theta _{\lbrack cd]}[(T_{\mu }^{[ac]}\eta
^{bd}-T_{\mu }^{[ad]}\eta ^{bc})-(T_{\mu }^{[bc]}\eta ^{ad}-T_{\mu
}^{[bd]}\eta ^{ac})]-\frac{1}{g}\partial _{\mu }\theta ^{\lbrack ab]}
\label{123a}
\end{equation}%
behaving as a two-index antisymmetric tensor field in the local frame and a
four-vector field in global spacetime.

The tensor field\ $T_{\mu \lbrack ab]}$ may in principle propagate, while
the tetrad $e^{\mu }$ is not yet treated as a dynamical field. The invariant
Lagrangian built from the tensor field strength
\begin{equation}
T_{\mu \nu }^{[ab]}=\partial _{\lbrack \nu }T_{\mu ]}^{[ab]}+g\eta
_{cd}T_{[\mu }^{[ac]}T_{\nu ]}^{[bd]}  \label{137}
\end{equation}%
can be written in a conventional form%
\begin{equation}
e\mathcal{L}_{G}=\frac{1}{2\kappa }e_{[a}^{\mu }e_{b]}^{\nu }T_{\mu \nu
}^{[ab]}\text{ , \ \ }e\equiv \lbrack -\det \mathrm{Tr}(e^{\mu }e^{\nu
})/4]^{-1/2}\text{ }  \label{127}
\end{equation}%
where $\kappa $\ stands for the modified Newtonian constant $8\pi /M_{Pl}^{2}
$. This form arises after using the commutator for tetrads and the standard
relations for $\gamma $ matrices\footnote{%
Some of these relations used throughout the paper are given below:%
\begin{eqnarray*}
\gamma ^{ab} &=&i[\gamma ^{a},\gamma ^{b}]/2,\text{ }\gamma ^{a}\gamma
^{b}=\gamma ^{ab}/i+\eta ^{ab}\mathbf{1},\text{ }\gamma _{c}\gamma ^{\lbrack
ab]}\gamma ^{c}=0 \\
\lbrack \gamma ^{ab},\gamma ^{a^{\prime }b^{\prime }}] &=&2i(\eta
^{ab^{\prime }}\gamma ^{ba^{\prime }}+\eta ^{ba^{\prime }}\gamma
^{ab^{\prime }}-\eta ^{aa^{\prime }}\gamma ^{bb^{\prime }}-\eta ^{bb^{\prime
}}\gamma ^{aa^{\prime }}) \\
Tr(\gamma ^{ab}\gamma ^{a^{\prime }b^{\prime }}) &=&4(\eta ^{aa^{\prime
}}\eta ^{bb^{\prime }}-\eta ^{ab^{\prime }}\eta ^{ba^{\prime }}),\text{ \ }%
Tr(\gamma ^{ab}\gamma _{cd})=4(\delta _{c}^{a}\delta _{d}^{b}-\delta
_{c}^{b}\delta _{d}^{a}) \\
Tr(\gamma ^{ab}\gamma ^{a^{\prime }b^{\prime }}\gamma ^{a^{\prime \prime
}b^{\prime \prime }}) &=&4i[\eta ^{aa^{\prime }}(\eta ^{a^{\prime \prime
}b^{\prime }}\eta ^{bb^{\prime \prime }}-\eta ^{a^{\prime \prime }b}\eta
^{b^{\prime }b^{\prime \prime }})+\eta ^{ab^{\prime }}(\eta ^{a^{\prime
\prime }b}\eta ^{a^{\prime }b^{\prime \prime }}-\eta ^{a^{\prime }a^{\prime
\prime }}\eta ^{bb^{\prime \prime }}) \\
&&+\eta ^{a^{\prime }b}(\eta ^{aa^{\prime \prime }}\eta ^{b^{\prime
}b^{\prime \prime }}-\eta ^{a^{\prime \prime }b^{\prime }}\eta ^{ab^{\prime
\prime }})+\eta ^{bb^{\prime }}(\eta ^{a^{\prime }a^{\prime \prime }}\eta
^{ab^{\prime \prime }}-\eta ^{a^{\prime }b^{\prime \prime }}\eta
^{aa^{\prime \prime }})]
\end{eqnarray*}%
where $\mathbf{1}$ in the above is the $4\times 4$ unit matrix.}. In fact,
this is the simplest pure gravity Lagrangian in the Palatini-type
formulation. Its variation with respect to the tensor field imposes a
constraint that expresses the tensor field in terms of the tetrads and their
derivatives, reducing $e\mathcal{L}_{G}$ to the standard Einstein
Lagrangian. The factor $e$, while not relevant for $SL(2,C)$ gauge
invariance, introduces an extra invariance of the action under general
four-coordinate transformations of $GL(4,R)$ \cite{ish}. The background
spacetime of the $SL(2,C)$ gravity is indeed the four-dimensional flat
Minkowski one. The $SL(2,C)$ does not alter the global spacetime structure
but instead defines the local symmetry group acting on tangent space or
spinor bundle. This is analogous to Yang-Mills theories, where gauge groups
act internally on fiber bundle over a spacetime manifold.

Interestingly, the tetrad invertibility conditions (\ref{or}), or more
precisely, their contracted form%
\begin{equation}
\frac{1}{4}e_{\mu }^{a}e_{a}^{\mu }=1  \label{1/4}
\end{equation}%
$\,$can be formally interpreted as the nonlinear $\sigma $-model type length
constraint that may lead to the condensation of tetrads. They can be
parametrized as an appropriate sum of symmetric and antisymmetric parts
\begin{subequations}
\begin{equation}
\text{\ }e_{\mu }^{a}\,=\eta _{\mu b}\left( \eta ^{ab}+\widetilde{e}%
^{\{ab\}}+\widehat{e}^{[ab]}\right) \,  \label{bd}
\end{equation}%
with the vacuum expectation value for the matrix-valued tetrad $e_{\mu }$
\end{subequations}
\begin{equation}
\overline{e}_{\mu }=\delta _{\mu }^{a}\gamma _{a}  \label{vev}
\end{equation}%
which represents an extremum of the action with $\delta _{\mu }^{a}$ and $%
\eta _{\mu b}\widehat{e}^{[ab]}$ acting as the effective Higgs mode and zero
modes, respectively.

This could be regarded as an ordinary example of spontaneous symmetry
breaking, if tetrads were treated as dynamical fields in the $SL(2,C)$ gauge
theory. As a consequence, six Goldstone-like zero modes $\widehat{e}^{[ab]}$
corresponding to the six broken $SL(2,C)$ generators would be expected.
However, in the absence of tetrad kinetic-energy terms in the gravity
Lagrangian (\ref{127}), these modes can be rotated away, rendering the
spontaneous symmetry breaking of $SL(2,C)$ effectively spurious.
Nevertheless, even when the tetrads are promoted to dynamical fields---as
considered later within the $SL(2N,C)$ framework---the spontaneous breaking
of the local frame $SL(2,C)$ symmetry does not affect the overall Poincar%
\'{e} invariance of the theory.

In the presence of fermions, the gauge-invariant matter coupling, expressed
via the covariant derivative (\ref{124}, \ref{125}) leads to an additional
interaction between the tensor field and the spin-current density
\begin{equation}
e\mathcal{L}_{M}=-\frac{1}{2}g\epsilon ^{abcd}T_{\mu \lbrack ab]}e_{c}^{\mu }%
\overline{\Psi }\gamma _{d}\gamma _{5}\Psi  \label{138}
\end{equation}%
This is a key feature of the Einstein-Cartan type gravity \cite{Kibble}
resulting, beyond standard General Relativity, in the tiny four-fermion
(spin current-current) interaction in the matter sector
\begin{equation}
\kappa \left( \overline{\Psi }\gamma _{d}\gamma ^{5}\Psi \right) (\overline{%
\Psi }\gamma ^{d}\gamma ^{5}\Psi )  \label{4f}
\end{equation}

\section{ Extending gravity: $\ SL(2N,C)$ gauge theories}

The view that the gravity governs the symmetry structure underlying the
hyperunification of all elementary forces necessitates extending the gauge
gravity symmetry group $SL(2,C)$ to the broader local symmetry $SL(2N,C)$,
where $N$ defines the degree of the internal $SU(N)$ symmetry as a subgroup.
We refer to this $SU(N)$ as hyperflavor symmetry, which incorporates all
known quantum numbers associated with quarks and leptons -- such as color,
weak isospin, and family numbers. Indeed, this hyperflavor $SU(N)$ symmetry
underpins the grand unification of the three other fundamental forces acting
on quarks and leptons. By proceeding along these lines, we aim to explore
the new insights that may arise from such a formulation.

\subsection{Basics of $SL(2N,C)$}

As noted above, the $SL(2N,C)$ symmetry group encompasses, among its primary
subgroups, the gravity $SL(2,C)$ symmetry --- which covers the orthochronous
Lorentz group --- and the internal hyperflavor $SU(N)$ symmetry proposed to
unify all quarks and leptons. Indeed, the $8N^{2}-2$ generators of $SL(2N,C)$
(which, as previously mentioned$^{1}$, can be viewed as a complexified
extension of the unitary $SU(2N)$ symmetry) are constructed from the tensor
products of the generators of $SL(2,C)$ and generators of $SU(N)$.
Consequently, the basic transformation applied to fermions takes the form
\begin{equation}
\Omega =\exp \left\{ \frac{i}{2}\left[ \left( \theta ^{k}+\theta
_{5}^{k}\gamma _{5}\right) \lambda ^{k}+\frac{1}{2}\theta _{ab}^{K}\gamma
^{ab}\lambda ^{K}\right] \right\} \text{ \ \ }(K=0,k)  \label{rt}
\end{equation}%
Among the $\lambda ^{K}$ matrices, $\lambda ^{k}$ $(k=1,...,N^{2}-1$)
represent the $SU(N)$\ Gell-Mann matrices, while $\lambda ^{0}=\sqrt{2/N}%
\mathbf{1}_{N}$ corresponds to the $U(1)$ generator involved (the $\theta $
parameters may either be constant or, in general, functions of spacetime
coordinate). Henceforth, we use uppercase Latin letters ($I,J,K$) for the $%
U(1)\times SU(N)$ symmetry, while the lowercase letters ($i,j,k$) for the
case of pure $SU(N)$ symmetry\footnote{%
The $\lambda $ matrices are normalized to satisfy%
\begin{eqnarray*}
\lbrack \lambda ^{k},\lambda ^{l}] &=&2if^{klm}\lambda ^{m},\text{ }%
\{\lambda ^{k},\lambda ^{l}\} \\
&=&\frac{4}{N}\delta ^{kl}\widehat{1}+2d^{klm}\lambda ^{m}),\text{ }%
Tr(\lambda ^{K}\lambda ^{L})=2\delta ^{KL}
\end{eqnarray*}%
}.

For description of the fermion matter in the theory one needs to reintroduce
the tetrad multiplet which generally has a form
\begin{equation}
e_{\mu }=(e_{\mu }^{aK}\gamma _{a}+e_{\mu 5}^{aK}\gamma _{a}\gamma
_{5})\lambda ^{K}\text{ \ }  \label{lllll}
\end{equation}%
It transforms, as before, according to (\ref{trl}), though the
transformation matrix is now given by equation (\ref{rt}). This tetrad
structure can be simplified by excluding its axial-vector component, which
can be achieved by imposing gauge-invariant constraints on the tetrads. To
facilitate this, a special nondynamical $SL(2N,C)$ scalar multiplet can be
introduced into the theory.
\begin{equation}
S=\exp \{i[(s^{k}+p^{k}\gamma _{5})\lambda ^{k}+t_{ab}^{K}\gamma
^{ab}\lambda ^{K}/2]\}  \label{ss}
\end{equation}%
which transforms like as $S\rightarrow \Omega S$. With this scalar multiplet
one can form a new tetrad in terms of the gauge invariant construction, $%
S^{-1}eS$. By appropriately choosing the flat-space components of the $S$
field, it becomes possible to eliminate the axial part of the tetrad \cite%
{ish1}, thereby reducing its final structure to
\begin{equation}
e_{\mu }=e_{\mu }^{aK}\gamma _{a}  \label{fs}
\end{equation}%
which we adopt in what follows.

\subsection{Gauging $SL(2N,C)$}

Once the $SL(2N,C)$ transformation (\ref{rt}) becomes local one needs, as
usual, to introduce the gauge field multiplet $I_{\mu }$\ transforming as
\begin{equation}
I_{\mu }\rightarrow \Omega I_{\mu }\Omega ^{-1}-\frac{1}{ig}(\partial _{\mu
}\Omega )\Omega ^{-1}  \label{gg}
\end{equation}%
whose strength-tensor takes the form
\begin{equation}
I_{\mu \nu }=\partial _{\lbrack \mu }I_{\nu ]}+ig[I_{\mu },I_{\nu }]
\label{i}
\end{equation}%
This provides the fermion multiplet with the covariant derivative
\begin{equation}
\partial _{\mu }\Psi \rightarrow D_{\mu }\Psi =\partial _{\mu }\Psi
+igI_{\mu }\Psi \text{ }  \label{cov}
\end{equation}%
where $g$ is the universal gauge coupling constant of the proposed
hyperunification. The $I_{\mu }$ multiplet generally consist of the vector,
axial-vector and tensor field submultiplets of $SU(N)$, and also the singlet
tensor field
\begin{equation}
I_{\mu }=V_{\mu }+A_{\mu }+T_{\mu }=\frac{1}{2}\left( V_{\mu }^{k}+A_{\mu
}^{k}\gamma _{5}\right) \lambda ^{k}+\frac{1}{4}T_{\mu \lbrack
ab]}^{K}\gamma ^{ab}\lambda ^{K}\text{ \ \ }(K=0,k)  \label{ggggg}
\end{equation}%
as follows from its decomposition into global spacetime components \footnote{%
In fact, all these submultiplets transform as four-vectors in global
spacetime. Meanwhile, in the local frame, the $V$ and $A$ submultiplets
behave are scalar and pseudoscalar fields, respectively, while the $T$
submultiplet transforms as a two-index antisymmetric tensor.}. Accordingly,
the corresponding total strength tensor written in the component fields
comes to
\begin{eqnarray}
I_{\mu \nu } &=&\frac{1}{2}\partial _{\lbrack \mu }\left( V^{k}+A^{k}\gamma
_{5}\right) _{\nu ]}\lambda ^{k}-\frac{1}{2}f^{ijk}g\left( V^{i}+A^{i}\gamma
_{5}\right) _{\mu }(V^{j}+A^{j}\gamma _{5})_{\nu }\lambda ^{k}  \notag \\
&&+\frac{1}{4}\left( \partial _{\lbrack \mu }T_{\nu ]}^{[ab]K}\gamma
_{ab}\lambda ^{K}+i\frac{g}{4}T_{\mu }^{[ab]K}T_{\nu }^{[a^{\prime
}b^{\prime }]K^{\prime }}[\lambda ^{K}\gamma _{ab},\lambda ^{K^{\prime
}}\gamma _{a^{\prime }b^{\prime }}]\right)   \label{f}
\end{eqnarray}%
Similarly, the gauge invariant fermion matter couplings, when given in terms
of the $I_{\mu }$\ submultiplets, take the form%
\begin{equation}
e\mathcal{L}_{M}=-\frac{g}{2}\overline{\Psi }\left\{ e^{\mu },\left[ \frac{1%
}{2}\left( V_{\mu }^{k}+A_{\mu }^{k}\gamma _{5}\right) \lambda ^{k}+\frac{1}{%
4}T_{\mu \lbrack ab]}^{K}\gamma ^{ab}\lambda ^{K}\right] \right\} \Psi
\label{lfm}
\end{equation}%
As one can readily observe, the vector, axial-vector and tensor fields
interact everywhere with the universal gauge coupling constant $g$ of the $%
SL(2N,C)$ HUT, which, in general, includes the conventional quadratic
strength terms for all gauge field submultiplets involved, alongside the
standard linear curvature Lagrangian for gravity (\ref{127}).\ \

This raises, as mentioned, a crucial question: how to ensure that only the
hyperflavored submultiplet of vector fields\ $V_{\mu }^{k}$ and a singlet
tensor field $T_{\mu \lbrack ab]}^{0}$ (underlying the $SU(N)$\ GUTs and
Einstein-Cartan gravity, respectively) appear in the observed particle
spectrum, while the hyperflavored submultiplets of axial-vector and tensor
fields, $A_{\mu }^{k}$ and $T_{\mu \lbrack ab]}^{k}$, are discriminated in
the theory. Remarkably, as shown later, this appears if the tetrads retain
their neutral pure-gravity form (\ref{tlf}), free of any $SU(N)$ components,
due to their constraining or condensation.

\subsection{Tetrads linked to gravity}

In fact, some challenge lies in the fact that the tetrads in (\ref{fs}),
along with a neutral component, also incorporate $SU(N)$ hyperflavored
components which may hinder the preservation of the tetrad invertibility
conditions in the extended theory. Consequently, this imposes a stringent
restriction on the permissible form of tetrads. To illustrate, let us assume
the tetrads adopt the general $SL(2N,C)$ covariant form
\begin{equation}
e_{\mu }=e_{\mu }^{aK}\gamma _{a}\lambda ^{K},\text{ }e_{\mu
}^{aK}e_{b}^{\mu K^{\prime }}=\Delta _{b}^{aKK^{\prime }},\text{ }e_{\mu
}^{aK}e_{a}^{\nu K^{\prime }}=\Delta _{\mu }^{\nu KK^{\prime }}  \label{or1}
\end{equation}%
with certain yet unspecified constructions for $\Delta _{b}^{aKK^{\prime }}$
and $\Delta _{\mu }^{\nu KK^{\prime }}$, which in the pure gravity case are
expected to satisfy the standard arrangement
\begin{equation}
\Delta _{b}^{a00}=\delta _{b}^{a}\text{ , \ }\Delta _{\mu }^{\nu 00}=\delta
_{\mu }^{\nu }  \label{or2}
\end{equation}%
Then multiplying the conditions (\ref{or1}) by the tetrad multiplets $%
e_{\sigma }^{bK^{\prime \prime }}$and $e_{a}^{\sigma K^{\prime \prime }}$,
respectively, one come after simple calculations to
\begin{equation}
\text{\ }e_{\mu }^{aK}=\Delta _{b}^{aK0}e_{\mu }^{b0}\text{, \ }e_{a}^{\mu
K^{\prime }}=\Delta _{a}^{bK^{\prime }0}e_{b}^{\mu 0}  \label{or3}
\end{equation}%
that finally gives
\begin{equation}
\Delta _{b}^{aKK^{\prime }}=\Delta _{c}^{aK0}\Delta _{b}^{cK^{\prime }0}
\label{or4}
\end{equation}%
and correspondingly
\begin{equation}
\Delta _{\mu }^{\nu KK^{\prime }}=\Delta _{\sigma }^{\nu K0}\Delta _{\mu
}^{\sigma K^{\prime }0}  \label{or4'}
\end{equation}%
For the constant and multiplicative forms of these functions, the only
viable solution arises as%
\begin{equation}
e_{\mu }^{aK}e_{b}^{\mu K^{\prime }}=\Delta _{b}^{aKK^{\prime }}=\delta
_{b}^{a}\delta ^{K0}\delta ^{K^{\prime }0}\text{ , \ }e_{\mu
}^{aK}e_{a}^{\nu K^{\prime }}=\Delta _{\mu }^{\nu KK^{\prime }}=\delta _{\mu
}^{\nu }\delta ^{K0}\delta ^{K^{\prime }0}  \label{or5}
\end{equation}%
which essentially mirrors the pure gravity case. Therefore, the
invertibility condition for tetrads holds only if they predominantly belong
to the $SL(2,C)$ subgroup rather than the entire $SL(2N,C)$ group
\begin{equation}
e_{\mu }^{aK}=e_{\mu }^{a}\delta ^{K0}  \label{or6}
\end{equation}%
Otherwise, their connection to general relativity is lost. Actually, just
the single neutral tetrad of $SU(N)$ appears consistent with basic tangent
bundle structure.

This can be achieved by directly imposing the constraint (\ref{or6}) on \
tetrads. Alternatively, when tetrads are treated as dynamical fields,
neutral tetrads may emerge via the spontaneous breaking of $SL(2N,C)$. This
way, inspired by the nonlinear $\sigma $-model type length constraint
(previously discussed in the pure gravity case)%
\begin{equation}
\frac{1}{4}e_{\mu }^{aK}e_{a}^{\mu K}=1  \label{n}
\end{equation}%
implies---when applied to general tetrads---that the initial symmetry can
break to $SL(2,C)\times SU(N)$. To provide such a vacuum configuration,
tetrads must take the form of an appropriate sum of symmetric and
antisymmetric parts%
\begin{equation}
\text{\ }e_{\mu }^{aK}\,=\eta _{\mu b}\left[ \,\eta ^{ab}\delta ^{K0}+%
\widetilde{e}^{\{ab\}K}+\widehat{e}^{[ab]K}\right]   \label{mai1}
\end{equation}%
with the vacuum expectation value for the matrix-valued tetrad $e_{\mu }$
\begin{equation}
\overline{e}_{\mu }=\delta _{\mu }^{a}\gamma _{a}\delta ^{K0}  \label{vev1}
\end{equation}%
which represents an extremum of the action with $\delta _{\mu }^{a}\delta
^{K0}$ and $\eta _{\mu b}\widehat{e}^{[ab]K}$ acting as the effective Higgs
mode and zero modes, respectively. Absorption of these modes induces, as we
see later, masses for the hyperflavored submultiplets of axial-vector and
tensor fields, $A_{\mu }^{k}$ and $T_{\mu \lbrack ab]}^{k}$, associated with
the broken coset generators beyond the residual $SL(2,C)\times SU(N)$
symmetry.

\section{Tetrad condensation scenario}

We consider the full $SL(2N,C)$ theory with the constraint (\ref{or6})
imposed on a general tetrad. This constraint is not a gauge fixing: local
gauge freedom is insufficient to align the internal index $K$ to the same
direction ($K=0$) at all spacetime points. Rather, it amounts either to a
fixed-background ansatz or to spontaneous symmetry breaking in the
dynamical-tetrad case, which we analyze below in this section.

\subsection{Constraints by direct tetrad filtering}

The essential point in considering the $SL(2N,C)$ gauge theory is that it
introduces an excessive number of degrees of freedom, some of which appear
irrelevant to observations. The key idea is that these degrees could be
dynamically suppressed due to the neutral tetrad (\ref{or6}) employed. Being
invertible and genuinely linked to Einstein-Cartan gauge gravity, this
tetrad is assumed not only to determine the geometric structure of spacetime
but also to regulate which local internal symmetries and associated gauge
field interactions are significant in the theory. Consequently, while the
full gauge multiplet of $SL(2N,C)$ typically includes vector, axial-vector,
and tensor field submultiplets of $SU(N)$, only the vector submultiplet and
the singlet tensor field can persist in the physical spectrum. This
effectively reduces the effective symmetry\ in the theory to $SL(2,C)\times
SU(N)$, thereby collecting together $SL(2,C)$ gauge gravity and $SU(N)$
grand unification, while all other gauge submultiplets are properly filtered
out by the tetrad.

Specifically, we proposed earlier \cite{jpl} that the actual gauge multiplet
$I_{\mu }$ (\ref{ggggg}) arises as result of the tetrad filtering of some
"prototype" nondynamical multiplet $\mathcal{I}_{\mu }$ which globally
transforms similarly to $I_{\mu }$, but, unlike it, does not itself gauge
the corresponding fermion system. These two multiplets are connected through
the tetrads via $I_{\mu }=(N/8)e_{\sigma }\mathcal{I}_{\mu }e^{\sigma }$,
introduced by the Lagrange multiplier-type term in the action. Consequently,
the gauge multiplet $I_{\mu }$ retains only those components of the
prototype multiplet $\mathcal{I}_{\mu }$, which survive the tetrad
filtering. In other words, the prototype multiplet $\mathcal{I}_{\mu }$
becomes partially dynamical to the extent permitted by the tetrads, which,
as shown, must be in the singlet-hyperflavor form (\ref{or6}) to preserve
the link to gravity.

However, with strictly invertible tetrads, such filtering eliminates the
gravity-inducing tensor fields from the gauge multiplet $I_{\mu }$ leaving
solely the submultiplets of vector and axial-vector fields. Only when the
tetrad invertibility condition is slightly relaxed, the appropriately
weakened tensor fields come into play. This occurs in a way that their
interaction essentially decouples from other elementary forces and
effectively adheres to the Einstein-Cartan type gravity action. The
corresponding curvature-squared terms constructed from the filtered tensor
fields appear to be vanishingly small and can be disregarded compared to the
standard strength-squared terms for vector fields. As a result, the entire
theory effectively exhibits a local $SL(2,C)\times SU(N)$ symmetry rather
than the hyperunified $SL(2N,C)$. While this approach appears successful in
many respects, it cannot be considered completely satisfactory, as it also
modifies the metric tensor, leading to a slight deviation from general
covariance, albeit in a controllable manner governed by a tiny parameter
related to the non-orthonormality of tetrads.

At the same time, the direct tetrad filtering with the hyperflavor-singlet
tetrads should not be regarded as a covariant construction in $SL(2N,C)$,
but rather as an illustrative preamble to its spontaneous symmetry breaking
through the tetrad condensation.

\subsection{Tetrad condensation}

Treating tetrads as dynamical fields, we add their invariant Lagrangian
taken in the canonical form
\begin{equation}
\mathcal{L}_{e}=-\frac{1}{4}\,(D_{\nu }e_{\mu }^{\;aK})(D^{\nu }e^{\mu
}{}_{aK})  \label{le}
\end{equation}%
including the covariant derivatives with an entire spin-connection multiplet
(\ref{ggggg}), to the total $SL(2N,C)$ Lagrangian. After the tetrad mass
rescaling
\begin{equation}
e_{\mu }^{\;aK}=M_{\!}\,E_{\mu }^{\;aK}  \label{rs}
\end{equation}%
where $E_{\mu }^{\;aK}$ stands for the dimensionless tetrad, one has for the
nonlinear $\sigma $-model type length constraint%
\begin{equation}
\frac{1}{4}\,e_{\mu }^{\;aK}e^{\mu K}{}_{a}=M_{\!}^{2}  \label{nc}
\end{equation}%
This, as mentioned above, leads to the spontaneous $SL(2N,C)$ symmetry
breaking being driven by the tetrad itself developing the VEV
\begin{equation}
\overline{e}_{\mu }^{\;}=M_{\!}\,\delta _{\mu }^{\;a}\,\gamma _{a}\,\lambda
^{0},\qquad \lambda ^{0}=\sqrt{2/N}\mathbf{1}_{N},  \label{e}
\end{equation}%
Because this VEV commutes with every generator $\lambda ^{k}$ but not with $%
\gamma _{5}\lambda ^{k}$ nor with $\gamma _{ab}\lambda ^{k}$, the vector
fields $V_{\mu }^{k}$ remain massless, while axial-vector fields $A_{\mu
}^{k}$ and all tensor fields $T_{\mu }^{ab\,K}$ ($K=0,k$) obtain masses of
order $gM_{\!}$ ($g$ is the hyperunified gauge coupling constant).
Remarkably, the $SL(2,C)$ symmetry survives being realized nonlinearly below
$M$; its six generators close on the $sl(2,C)$ algebra even though they lie
in the coset. Thereby, an actual residual symmetry is, in fact, $%
SL(2,C)\times SU(N),$ solely associated with the vector field multiplet in
the internal space.

This framework introduces two fundamental mass scales -- the Planck scale $%
M_{\!P}$ and the scale $M$ where $SL(2N,C)$ breaks. Above $M$, the
spin-connection fields propagate with standard Yang--Mills--type kinetic
terms. Below the scale $M$ the action contains
\begin{equation}
\frac{M_{P}^{2}}{2}\,R(e)+\frac{1}{4}(De)^{2}+\frac{c_{R}}{M^{2}}%
\,R^{2}+\ldots   \label{mp}
\end{equation}%
including the quadratic curvature terms. The algebraic equation of motion
for the heavy spin connection enforces the metricity condition, $D_{\mu
}e_{\nu }^{\;a}=0$, up to $O(p^{2}/M^{2})$, so varying with respect to the
tetrad $e_{\mu }^{\;a}$ (rather its symmetrical part (\ref{bd})) yields the
standard Einstein equation plus Planck-suppressed curvature/torsion
corrections. Classical tests of GR remain unaffected.

The point is, however, the constraint (\ref{nc}) itself only fixes the
length of the unified tetrad but leaves a continuous family of VEV
orientations. Another problem is that in a general tetrad (\ref{mai1}),
while asymmetrical components $\widehat{e}^{[ab]K}$are eaten by the gauge
axial-vector and tensor field multiplets, the symmetrical components $%
\widetilde{e}^{\{ab\}K}$ are left in the theory. They carry the $SU(N)$
hyperflavor and can become serious problem unless acquire heavy masses.
Fortunately, quantum gauge fluctuations around a background $\bar{e}_{\mu }$
generate a Coleman--Weinberg (CW) potential \cite{cw}, which, as appears,
provides solution to both issues. It lifts degeneracy in favor of the
"neutral" vacuum (\ref{e}), and ensures heavy masses for the hyperflavored
tetrad fluctuations.

\subsection{Hyperflavor-blind vacuum}

The tetrad length constraint is imposed by a Lagrange multiplier,
\begin{equation}
\frac{1}{32}\mathrm{Tr}(e_{\mu }e^{\mu })=M^{2}\text{ \ \ \ \ }  \label{eql}
\end{equation}%
which fixes the norm but not the orientation in the internal (hyperflavor)
space.

A convenient vacuum parametrization is
\begin{equation}
\bar{e}_{\mu }\;=\;v_{\mu }\otimes H,\qquad v_{\mu }=M\,\delta _{\mu
}^{\,a}\gamma _{a},\qquad H^{\dagger }=H,  \label{eqv}
\end{equation}%
so that all internal-orientation dependence sits in the $N\times N$ matrix $H
$. The "hyperflavor-blind" choice $H\propto \mathbf{1}_{N}$, corresponding
to the the residual $SU(N)$ symmetry, is one of the possible vacua that
satisfies the constraint (\ref{eql}). However, radiative corrections render
it a strict minimum. Notably, just the vector field multiplet piece of the
CW potential turns out orientation-selective. The axial-vector and tensor CW
pieces are insignificant for the vacuum orientation. Likewise, fermion loops
neither favor nor disfavor the HB vacuum---they add only an
orientation-independent constant.

The one-loop CW potential from vector fields in Landau gauge is
\begin{equation}
U_{V}(e)\;=\;\frac{3}{64\pi ^{2}}\sum_{n}m_{n}^{4}(e)\left( \ln \frac{%
m_{n}^{2}(e)}{\mu ^{2}}-\frac{5}{6}\right) ,  \label{eqc}
\end{equation}%
where $m_{n}^{2}$ are the eigenvalues of the gauge-boson mass matrix $%
M_{V}^{2}(e)$ and $\mu $ is the renormalization scale. For gauge generators $%
T_{V}=\lambda _{k}$ of the $SU(N)$ symmetry group, the vector field mass
operator in the background $\bar{e}$ is given by
\begin{eqnarray}
(M_{V}^{2})_{kl}(\bar{e})\; &=&\;g^{2}\mathrm{Tr}\!\left( [\lambda _{k},\bar{%
e}_{\mu }][\lambda _{l},\bar{e}^{\mu }]\right)  \label{eqm} \\
&=&4g^{2}M^{2}\,\mathrm{Tr}\!\left( [\!\lambda _{k},H][\!\lambda
_{l},H]\right)  \notag
\end{eqnarray}%
where we have used (\ref{eqv}),
\begin{equation}
\lbrack \lambda _{k},\bar{e}_{\mu }]\;=\;v_{\mu }\otimes \lbrack \,\lambda
_{k},H\,]  \label{te}
\end{equation}

As one can readily see, the potential $U_{V}$ is minimized uniquely at the
hyperflavor-blind orientation $H\propto \mathbf{1}_{N}$, where it vanishes.
If $H$ is not proportional to $\mathbf{1}_{N}$, then some eigenvalues $%
m_{n}^{2}$ of the mass matrix $M^{2}(\bar{e})$ are non-zero and, thereby,
the potential $U_{V}>0$. Thus, within the vector field sector, the
hyperflavor-blind orientation is a strict minimum. Any non-HBV spectrum
lifts it higher. The axial-vector and tensor CW pieces, $U_{A,T}(\bar{e})$,
are orientation-blind at HBV: their masses $\sim gM$ depend on the Dirac
algebra, not on $H$ near $H\propto \mathbf{1}_{N}$, so they contribute a
constant and do not tilt the internal orientation. However, axial/tensor
loops do curve the potential along symmetric tetrad directions, as is seen
below.

Consider this in some detail. A real tetrad $4\times 4$ matrix has $16$
components whose splitting into Lorentz--antisymmetric and symmetric parts
gives $6$ and $10$ per internal index $K$, thus giving $16N^{2}$ components
altogether. At the HBV, spontaneous breaking by the singlet tetrad VEV leads
to the following \textquotedblleft eating\textquotedblright\ pattern: all
the $6N^{2}$ antisymmetric (torsion-linked) tetrad modes are eaten by the
tensor gauge fields $T_{\mu }^{abK}$, while among the $10N^{2}$ symmetric
modes, the $N^{2}\!-\!1$ adjoint traces $\widetilde{e}_{\mu }^{\mu k}{}$ are
eaten by the axial vectors $A_{\mu }^{k}$ (with $k=1,\dots ,N^{2}\!-\!1$).
Among the remaining $9N^{2}+1$ symmetric modes the $9(N^{2}\!-\!1)$
components are $SU(N)$-adjoint symmetric traceless modes, while the rest $%
SU(N)$-singlet\emph{\ }$10$ components split into $9$ traceless spin-2
components linked to graviton (two physical helicities after
gauge/constraints) and one trace. The trace is fixed by the length
constraint and does not propagate. At the HBV the $SU(N)$-adjoint symmetric
modes acquire Coleman--Weinberg masses from axial-vector and tensor loops,
while the genuine hyperflavor-singlet graviton remains massless.

Indeed, in the CW potential analogs for axial-vector and tensor multiplets, $%
U_{A}(e)$ and $U_{T}(e)$, the corresponding mass matrices have the forms
\begin{equation}
M_{A,T}^{2}(e)\;\propto \;g^{2}\,\mathrm{Tr}\!([T_{A,T},e_{\mu
}][T_{A,T},e^{\mu }]),\qquad T_{A}=\gamma _{5}\lambda ^{k},\;\;T_{T}=\gamma
_{ab}\lambda ^{K}  \label{eqm1}
\end{equation}%
respectively. Expanding $e_{\mu }=\overline{e}_{\mu }+\varepsilon _{\mu }$
around the HBV (\ref{eqv}) and keeping $O(\varepsilon ^{2})$, the
axial/tensor loops generate a positive-definite curvature along the adjoint
symmetric-traceless directions that produces masses for the hyperflavored
fluctuations $\varepsilon _{\mu }^{k}$ of the order $O(g^{2}M)$
\begin{equation}
\Delta \mathcal{L}_{\mathrm{CW}}^{(2)}\;=\;-\frac{1}{2}\varepsilon ^{k}\cdot
\mathcal{M}^{2}\cdot \varepsilon ^{k},\qquad \mathcal{M}^{2}\sim \frac{g^{4}%
}{16\pi ^{2}}\,M^{2}\,
\end{equation}%
across the $9(N^{2}\!-\!1)$ modes.

The graviton-linked components corresponds to the $SU(N)$-singlet member of
the symmetric traceless $9N^{2}\!+\!1$ set. For this fluctuation, $%
\varepsilon ^{0}\propto \lambda ^{0}$, we have
\begin{equation}
\lbrack \bar{e}_{\mu },\,\varepsilon ^{0}]=0\,
\end{equation}%
so neither the tree--level mixing with heavy fields nor the axial/tensor
loops induce a mass. Thus, as was expected, gravity part is completely
determined by the usual flavour-singlet tetrad%
\begin{equation}
e_{\mu }^{\;aK}(x)\;\longrightarrow \;e_{\mu }^{\;a}(x)\,\delta ^{K0},\text{
}e_{\mu }^{\;a}=\varepsilon _{\mu }^{a0}  \label{ar}
\end{equation}%
while all $K\neq 0$ tetrad components decouple from the particle spectrum
being eaten by the axial-vector and tensor multiplets (Lorentz-asymmetrical
states) or\ acquiring heavy masses from radiative corrections
(Lorentz-asymmetrical ones).

A few final remarks are in order:

($i$) The dynamical tetrad alone breaks all non-compact directions at some
scale $M$, leaving an unbroken compact $SU(N)$ GUT for the hyperflavor
sector;

($ii$) Local Lorentz symmetry survives being realized nonlinearly below $M$:
its six generators, while lying in the coset, still form $sl(2,C)$ algebra;

($iii$) Two scales in the theory --- $M_{P}$ sets Newton's constant, while a
freely-chosen $M$ (potentially as low as a few TeV) sets the heavy
axial-vector and tensor field multiplets threshold;

($iv$) Einstein--Cartan structure re-emerges with at most $1/M^{2}$
corrections; torsion effects set by a scale $M$.

One can see that the tetrad-condensation scenario not only complements the
direct tetrad-contraint approach, but also lends stronger justification to
its intuitive conclusions, rendering them more concrete and far-reaching in
scope.

\section{Lagrangians}

\subsection{Linear strength theories}

We are taking the linear hyperunified theory Lagrangian in the Palatini type
form
\begin{equation}
e\mathcal{L}_{H}^{(1)}\sim Tr\{[e^{\mu },e^{\nu }]I_{\mu \nu }\}
\label{127a}
\end{equation}%
In the $SL(2N,C)$ case the strength tensor $I_{\mu \nu }$ (\ref{f}), apart
from tensor submultiplet underlying the gravity sector, comprises the vector
and axial-vector submultiplets as well. However, due to the neutral tetrad (%
\ref{ar}) emerged in the symmetry broken phase, they, as follows from the
commutator,%
\begin{equation}
\lbrack e^{\mu },e^{\nu }]=-2ie_{a}^{\mu }e_{b}^{\nu }\gamma ^{ab}
\label{Tr2}
\end{equation}%
do not contribute to the linear Lagrangian (\ref{127a}).

Eventually, for the tensor field strength in (\ref{f}) one has, after taking
the traces of products involving $\gamma $ and $\lambda $ matrices, the
following linear curvature gravity Lagrangian
\begin{equation}
e\mathcal{L}_{H}^{(1)T}=\frac{1}{2\kappa }\left( \partial _{\lbrack \mu
}T_{\nu ]}^{[ab]0}+g\eta _{cd}T_{[\mu }^{[ac]K}T_{\nu ]}^{[bd]K}\right)
e_{a}^{\mu }e_{b}^{\nu }  \label{p2}
\end{equation}%
Similarly, there are also couplings with the tensor submultiplet, as derived
from the general matter Lagrangian (\ref{lfm})
\begin{equation}
e\mathcal{L}_{HM}^{T}=-\frac{g}{2}\epsilon ^{abcd}T_{\mu \lbrack ab]}^{K}%
\overline{\Psi }e_{c}^{\mu }\gamma _{d}\lambda ^{K}\gamma ^{5}\Psi
\label{m1}
\end{equation}%
where interactions of tensor fields with the neutral and hyperflavored spin
density currents appear with the same coupling constant $g$.

Notably, in the Lagrangian (\ref{p2}), only the singlet component $T_{\mu
}^{[ab]0}$of the total tensor field multiplet $T_{[\mu }^{[ab]K}$ contains
derivative terms, while the hyperflavored components $T_{\mu }^{[ab]k}$ are
solely involved through interaction terms. This indicates that the singlet
tensor field alone gauges gravity, while the hyperflavored components simply
convert to spin currents, given by $\epsilon ^{abcd}\overline{\Psi }e_{\mu
c}\gamma _{d}\lambda ^{k}\gamma ^{5}\Psi $. When they both, $T_{[\mu
}^{[ab]0}$ and $T_{\mu }^{[ab]k}$, are independently eliminated from the
entire linear tensor field Lagrangian $e\mathcal{L}_{H}^{(1)T}+e\mathcal{L}%
_{M}^{(T)}$, one arrives at the Einstein-Cartan type gravity containing,
besides the usual GR, the tiny four-fermion spin density interaction
\begin{equation}
\kappa \left( \overline{\Psi }\gamma _{c}\gamma ^{5}\lambda ^{K}\Psi \right)
(\overline{\Psi }\gamma ^{c}\gamma ^{5}\lambda ^{K}\Psi )  \label{6}
\end{equation}%
which in contrast to the standard case \cite{Kibble} includes the
hyperflavored four-fermion interaction terms as well.

\subsection{Quadratic strength theories}

\subsubsection{Vector and axial-vector fields}

Let us focus first of all on the vector and axial-vector fields which are
the basic spin-1 carriers of the hyperflavor $SU(N)$ symmetry in the $%
SL(2N,C)$ theory. Their own sector stemming from the common strength tensor (%
\ref{f}) looks as%
\begin{eqnarray}
e\mathcal{L}_{H}^{VA} &=&-\frac{1}{2}\mathrm{Tr}[(V_{\mu \nu }^{k}\lambda
^{k}+\gamma _{5}A_{\mu \nu }^{k}\lambda ^{k})^{2}/4]=-\frac{1}{4}(V_{\mu \nu
}^{k})^{2}-\frac{1}{4}(A_{\mu \nu }^{k})^{2}  \label{I} \\
&=&-\frac{1}{4}[\partial _{\lbrack \mu }V_{\nu ]}^{k}-gf^{ijk}(V_{\mu
}^{i}V_{\nu }^{j}+A_{\mu }^{i}A_{\nu }^{j})]^{2}-\frac{1}{4}[\partial
_{\lbrack \mu }A_{\nu ]}^{k}-gf^{ijk}(V_{\mu }^{i}A_{\nu }^{j})]^{2}  \notag
\end{eqnarray}%
where, as one can see, the vector fields acquire a conventional $SU(N)$
gauge theory form, while the axial-vector multiplet looks like as an adjoint
matter in this theory. However, some of its couplings break this gauge
invariance, as follows from the second line terms\footnote{%
Tensor field contributions to the field-strengths $V_{\mu \nu }^{k}$and $%
A_{\mu \nu }^{k}$, which likewise break this $SU(N)$ symmetry, are omitted
here. As shown in Section 4, the tensor fields acquire Planck-scale masses
and therefore decouple from low-energy dynamics.
\par
{}}. At the same time, in the matter sector of the theory (\ref{lfm}), the
vector fields interact with ordinary matter fermions%
\begin{equation}
e\mathcal{L}_{HM}^{V}=-\frac{g}{2}V_{\mu }^{i}\overline{\Psi }e_{a}^{\mu
}\gamma ^{a}\lambda ^{i}\Psi   \label{m3}
\end{equation}%
whereas the axial-vector fields remain completely decoupled due to the
tetrads involved.

Nonetheless, as indicated by the gauge sector Lagrangian (\ref{I}), they may
still participate in various processes mediated by trilinear and
quadrilinear couplings with vector fields. While current experimental data
\cite{pdg} do not definitively exclude such processes, direct evidence for
axial-vector fields remains lacking. Still, one may seek to reconcile their
existence with observations. As was shown above, they typically become
massive during the starting symmetry breaking to $SL(2,C)\times SU(N)$ and
may, in principle, be detectable, if located in the multi-TeV region.

Another scenario might be, if the axial-vector fields condense at some
Planck-scale order mass $\mathcal{M}$, thus providing a true vacuum in the
theory, $\left\langle A_{\mu }^{k}\right\rangle =\mathfrak{n}_{\mu }^{k}%
\mathcal{M}$, whose direction is given by the unit Lorentz vector $\mathfrak{%
n}_{\mu }^{k}$ ($\mathfrak{n}_{k}^{\mu }\mathfrak{n}_{\mu }^{k}=1$).
Remarkably, in such a vacuum, as is directly seen, gauge invariance for the
vector fields is fully restored, though a tiny spontaneous breaking of the
Lorentz invariance at the scale $\mathcal{M}$ may appear \cite{jpl}.

\subsubsection{Entire gauge multiplet}

As earlier noted, extending the $SL(2,C)$ gauge gravity to the hyperunified $%
SL(2N,C)$ theory naturally implies the inclusion of some "safe" quadratic
curvature terms in its gravitational sector, alongside the standard
quadratic terms for the vector field strength tensors. This criterion
distinctly identifies the ghost-free curvature-squared gravity model,
originally proposed by Neville \cite{nev}, as the most fitting for such an
extension.

Accordingly, we employ the corresponding Lagrangian.

\begin{equation}
e\mathcal{L}_{G}^{(2)}=\lambda T_{abcd}\left(
T^{abcd}-4T^{acbd}+T^{cdab}\right)  \label{r2'}
\end{equation}%
where the curvature tensor $T_{\mu \nu }^{ab}$ is contracted with tetrads, $%
T^{abcd}=T_{\mu \nu }^{ab}e^{\mu c}e^{\nu d}$, and properly
(anti)symmetrized. In the $SL(2,C)$ gravity case this curvature is
constructed from tensor fields involved, while the constant $\lambda $,
though currently arbitrary, may potentially be determined within the $%
SL(2N,C)$ unification scheme. The Lagrangian $\mathcal{L}_{G}^{(2)}$, when
properly expressed through the tensor field strengths and tetrad components
takes the form
\begin{equation}
e\mathcal{L}_{G}^{(2)}=\lambda (T_{ab}^{\mu \nu }T_{\mu \nu
}^{ab}-4T_{ab}^{\mu \nu }T_{\rho \nu }^{ac}e_{\mu c}e^{\rho b}+T_{ab}^{\mu
\nu }T_{\rho \sigma }^{cd}e_{\mu c}e_{\nu d}e^{\rho a}e^{\sigma b})
\label{re}
\end{equation}%
Consequently, the spin-connection field $T_{\mu }^{ab}$ becomes genuinely
dynamic. In the particle spectrum, alongside the ordinary massless graviton,
there exists a massive excitation which, in principle, could normally
propagate \cite{nev, nev1}. However, it typically possesses a Planck-scale
order mass, $M_{P}^{2}/\lambda $, making it unlikely to have any observable
significance unless the numerical parameter $\lambda $ is exceptionally
large.

In extending the $SL(2,C)$ to the hyperunified $SL(2N,C)$ framework, it is
natural to view all terms in the gravity Lagrangian (\ref{re}) as descending
from strength traces of the full $SL(2N,C)$ gauge multiplet in the quadratic
sector,
\begin{equation}
e\mathcal{L}_{H}^{(2)}=\lambda _{I}\mathrm{Tr}(aI^{\mu \nu }I_{\mu \nu
}+bI^{\mu \nu }I_{\rho \nu }e_{\mu }e^{\rho }+cI^{\mu \nu }I_{\rho \sigma
}e_{\mu }e_{\nu }e^{\rho }e^{\sigma })  \label{r1'}
\end{equation}%
which in the pure gravity limit comes to the $e\mathcal{L}_{G}^{(2)}$, and
the constant $a$, $b$ and $c$, still arbitrary, are to be fixed by this
correspondence.

Let us first focus on the vector and axial-vector field submultiplets in
this extension and examine how the internal symmetry in the hyperunified
theory is organized. This can be analyzed by considering the Lagrangian (\ref%
{r1'}) term by term using neutral tetrads (\ref{ar}) emerged from the
spontaneous $SL(2N,C)$ violation through the tetrad condensation. As
follows, we come to the "almost" gauge-invariant form given above in the
Lagrangian $e\mathcal{L}^{VA}$ (\ref{I}). Indeed, from the structure of the
vector and axial-vector field submultiplets in the general strength-tensor $%
I_{\mu \nu }$\ (\ref{f}) one obtains the following expressions for the
first, second, and third terms in (\ref{r1'}), respectively

\begin{eqnarray}
\mathrm{Tr}\left( I^{(W)\mu \nu }I_{\mu \nu }^{(W)}\right) &=&(W_{\mu \nu
}^{k}W^{\mu \nu k})/2,\text{ }  \notag \\
\mathrm{Tr}\left( I^{(W)\mu \nu }I_{\rho \nu }^{(W)}e_{\mu }e^{\rho }\right)
&=&2(W_{\mu \nu }^{k}W^{\mu \nu k}),  \notag \\
\mathrm{Tr}\left( I^{(W)\mu \nu }I_{\rho \sigma }^{(W)}e_{\mu }e_{\nu
}e^{\rho }e^{\sigma }\right) &=&4(W_{\mu \nu }^{k}W^{\mu \nu k})\text{ \ }%
(W\equiv V,A)  \label{r0}
\end{eqnarray}%
where the tetrad invertibility conditions (\ref{or5}) and corresponding
traces for $\gamma $ and $\lambda $ matrices have been applied.\ The
analogous terms for tensor field submultiplet have the forms
\begin{eqnarray}
\mathrm{Tr}\left( I^{(T)\mu \nu }I_{\mu \nu }^{(T)}\right) &=&T_{ab}^{\mu
\nu K}T_{\mu \nu }^{abK},\text{ }  \notag \\
\mathrm{Tr}\left( I^{(T)\mu \nu }I_{\rho \nu }^{(T)}e_{\mu }e^{\rho }\right)
&=&3T_{ab}^{\mu \nu K}T_{\mu \nu }^{abK}+4(T_{ab}^{\mu \nu K}T_{\rho \nu
}^{acK}e_{\mu c}e^{\rho b}),  \notag \\
\mathrm{Tr}\left( I^{(T)\mu \nu }I_{\rho \sigma }^{(T)}e_{\mu }e_{\nu
}e^{\rho }e^{\sigma }\right) &=&2T_{ab}^{\mu \nu K}T_{\mu \nu
}^{abK}+8(T_{ab}^{\mu \nu K}T_{\rho \sigma }^{cdK}e_{\mu c}e_{\nu d}e^{\rho
a}e^{\sigma b})\text{ }  \label{r00}
\end{eqnarray}%
\

These relations lead to the following gauge-invariant combination for the
vector, axial-vector, and tensor submultiplets in the general hyperunified
Lagrangian
\begin{equation}
(a/2+3b+4c)W^{2}+(a+3b+2c)TT+4bTTee+8cTTeeee  \label{W2}
\end{equation}%
This imply that, to match the ghost-free Lagrangian for tensor fields in the
pure-gravity case (\ref{re}), one must impose the conditions
\begin{equation}
a+3b+2c=1,\text{ }b=-1,\text{ }c=1/8\rightarrow a=15/4  \label{abc}
\end{equation}%
that fixes the constants in the general Lagrangian (\ref{r1'}), and also the
$W^{2}$ coefficient as $3/8$ in (\ref{W2}). Accordingly, the hyperunified $%
SL(2N,C)$ Lagrangian acquires the form
\begin{eqnarray}
e\mathcal{L}_{H}^{(2)} &=&-\frac{1}{4}(V_{\mu \nu }^{k}V^{\mu \nu k}+A_{\mu
\nu }^{k}A^{\mu \nu k})  \label{RR} \\
&&-\frac{1}{4}\left( T_{ab}^{\mu \nu K}T_{\mu \nu }^{abK}-4T_{ab}^{\mu \nu
K}T_{\rho \nu }^{acK}e_{\mu c}e^{\rho b}+T_{ab}^{\mu \nu K}T_{\rho \sigma
}^{cdK}e_{\mu c}e_{\nu d}e^{\rho a}e^{\sigma b}\right)   \notag
\end{eqnarray}%
where the overall factor $\lambda _{I}=-2/3$ is chosen to put the vector and
axial-vector sectors in canonical form. The tensor sector is brought to the
same canonical form by the following rescaling of tensor fields and their
coupling%
\begin{equation}
T_{ab}^{\mu \nu K}\rightarrow \sqrt{8/3}T_{ab}^{\mu \nu K},\text{ }%
g\rightarrow \sqrt{3/8}g  \label{gt}
\end{equation}%
so that a single gauge coupling $g$ governs the entire quadratic Lagrangian.
Consequently, we arrive at a consistent hyperunified scheme in which, aside
from the linear gravity Lagrangian (\ref{p2})---to which only tensor fields
contribute---the quadratic sector is fully unified with all gauge
submultiplets included.

Finally, combining the linear and quadratic curvature terms for the tensor
fields, as given in (\ref{p2}) and (\ref{RR}), respectively, one is lead to
a remarkable conclusion regarding the heavy mass origin of tensor fields. In
fact, this directly follows from their polynomial terms in (\ref{p2})
\begin{equation}
e\frac{1}{2\kappa }g\eta _{cd}\left( T_{\mu }^{[ac]K}T_{\nu }^{[bd]K}-T_{\nu
}^{[ac]K}T_{\mu }^{[bd]K}\right) e_{a}^{\mu }e_{b}^{\nu }\text{ \ \ \ \ }%
(\mu ,\nu ;c,d=0,1,2,3)  \label{lh1}
\end{equation}%
In flat spacetime, where $e_{a}^{\mu }=\delta _{a}^{\mu }$ and $e_{b}^{\nu
}=\delta _{b}^{\nu }$, these couplings yield a Fierz-Pauli type mass term
for all tensor fields
\begin{equation}
\frac{g}{2\kappa }\eta _{cd}\left( T_{\mu }^{[\mu c]K}T_{\nu }^{[\nu
d]K}-T_{\nu }^{[\mu c]K}T_{\mu }^{[\nu d]K}\right) \text{ \ }  \label{lh2}
\end{equation}%
which ultimately leads to the mass%
\begin{equation}
M_{T}^{2}\sim gM_{P}^{2}  \label{3}
\end{equation}%
This implies that, in addition to the mass generated by tetrad condensation
(previous section), the linear-strength Lagrangian (\ref{p2}) drives the
tensor field submultiplet $T_{\mu }^{[ac]K}$ to superheavy mass, decoupling
them from the gauge scale $M$ and pushing them toward the Planck scale $M_{P}
$. Consequently, only the axial-vector submultiplet remains at the scale $M$.

\section{Application to GUTs}

\subsection{Symmetry breaking scenario}

It is evident that the hyperunification of all elementary forces implies
that, while gravity is basically governed by its unique linear tensor field
strength Lagrangian (\ref{p2}), the quadratic strength terms for all
components of gauge multiplet $I_{\mu }$ are naturally unified in a common $%
SL(2N,C)$ invariant Lagrangian (\ref{RR}). After symmetry breaking caused by
the tetrad condensation, one comes to the effective $SL(2,C)\times SU(N)$
theory integrating both the Einstein-Cartan type gravity and the standard $%
SU(N)$ invariant vector field interactions within a single framework.
Independently, such a symmetry reduction to $SL(2,C)\times SU(N)$ appears in
the composite model of quarks and leptons, as we show later.

A conventional scenario for\ further breaking of the $SU(N)$ invariance in
the theory---considered as some GUT candidate---introduces a suitable set of
scalar fields that may reduce this theory down to the Standard Model. For
this, one takes the adjoint scalar multiplets of the type
\begin{equation}
\Phi =(\phi ^{k}+i\phi _{5}^{k}\gamma _{5})\lambda ^{k}+\phi _{ab}^{K}\gamma
^{ab}\lambda ^{K}/2  \label{ad}
\end{equation}%
which transform under $SL(2N,C)$ as%
\begin{equation}
\Phi \rightarrow \Omega \Phi \Omega ^{-1}\text{\ }  \label{fo}
\end{equation}%
In general, $\Phi $ contains not only scalar components but also
pseudoscalar and spinorial tensor components in (\ref{ad}). These can be
filter out by the covariant tetrad-constraint mechanism. Imposing on $\Phi $%
\ \ the constraint
\begin{equation}
\Phi =e_{\sigma }\Phi e^{\sigma }/4  \label{f1}
\end{equation}%
one finds, after tetrad condensation, that $\Phi $ is effectively
\textquotedblleft sandwiched\textquotedblright\ by the neutral tetrads\ (\ref%
{ar}). As a result, only the pure scalar components remain in the $SU(N)$
symmetry breaking multiplet $\Phi $
\begin{equation}
\Phi =\phi ^{k}\lambda ^{k}\text{ \ \ \ \ }(k=1,...,N^{2}-1)\text{\ \ }
\label{f2}
\end{equation}%
thereby enabling (together with similar scalar multiplets) the breaking of
the $SU(N)$ GUT down to the Standard Model. The final symmetry breaking to $%
SU(3)_{c}\times U(1)_{em}$ is achieved by additional scalar multiplets with
assignments determined by representations chosen for quarks and leptons.

\subsection{$SU(5)$ and its direct extensions}

As discussed, the $SL(2N,C)$ HUT effectively exhibits a local $SL(2,C)\times
\ SU(N)$ symmetry, rather than an entire $SL(2N,C)$ symmetry, which
primarily serves to determine the structure of the gauge and matter
multiplets. This results in the $SL(2,C)$ gauge gravity, on one side, and $%
SU(N)$ grand unified theory, on the other. Given that all states involved in
the $SL(2N,C)$ theories are additionally classified according to their spin
magnitudes, many potential $SU(N)$ GUTs, including the conventional $SU(5)$
theory \cite{gg}, appear to be irrelevant for the standard spin-$1/2$ quarks
and leptons. However, the application of $SL(2N,C)$ to their proposed preon
constituents turns out both natural and exceptionally promising.

Note first that the $SL(2N,C)$ symmetry is presently applied to the chiral
fermions, leading to the decomposition of a general transformation (\ref{rt}%
) into distinct transformations for left-handed and right-handed fermions
\begin{eqnarray}
\Omega _{L} &=&\exp \left\{ \frac{i}{2}\left[ \left( \theta ^{k}-i\theta
_{5}^{k}\right) \lambda ^{k}+\frac{1}{2}\theta _{ab}^{K}\Sigma ^{ab}\lambda
^{K}\right] \right\} ,\text{ \ }  \notag \\
\Omega _{R} &=&\exp \left\{ \frac{i}{2}\left[ \left( \theta ^{k}+i\theta
_{5}^{k}\right) \lambda ^{k}+\frac{1}{2}\theta _{ab}^{K}\overline{\Sigma }%
^{ab}\lambda ^{K}\right] \right\} \text{ \ }  \label{lr}
\end{eqnarray}%
in the chiral basis for\ $\gamma $\ matrices.\ Here $\Sigma ^{ab}$ and $%
\overline{\Sigma }^{ab}$ are given by
\begin{equation}
\Sigma ^{ab}=\frac{i}{2}(\sigma ^{a}\overline{\sigma }^{b}-\sigma ^{b}%
\overline{\sigma }^{a}),\text{ \ }\overline{\Sigma }^{ab}=\frac{i}{2}(%
\overline{\sigma }^{a}\sigma ^{b}-\overline{\sigma }^{b}\sigma ^{a})
\label{r1}
\end{equation}%
where the two-dimensional matrices $\sigma ^{a}$ and $\overline{\sigma }^{a}$
\begin{equation}
\gamma ^{a}=\left(
\begin{array}{cc}
0 & \sigma ^{a} \\
\overline{\sigma }^{a} & 0%
\end{array}%
\right)   \label{r3}
\end{equation}%
are expressed in terms of the unit and Pauli matrices as%
\begin{equation}
\sigma ^{a}=(1,\boldsymbol{\sigma }),\text{ }\overline{\sigma }^{a}=(1,-%
\boldsymbol{\sigma })  \label{r2}
\end{equation}%
Accordingly, the gauge multiplets of $SL(2N,C)$\ associated with both
left-handed and right-handed fermions are appropriately specified.

We begin with the familiar $SU(5)$, which could naturally emerge from the $%
SL(10,C)$ hyperunification. In this context, some of its low-dimensional
chiral fermion multiplets (left-handed for certainty) can be represented in
terms of the $SU(5)\times SL(2,C)$ components as
\begin{equation}
\Psi _{L}^{ia}\text{ },\text{\ \ \ }10=(\overline{5},2)\text{ }  \label{sl1}
\end{equation}%
and
\begin{equation}
\Psi _{L[ai,\text{ }jb]}=\Psi _{L[ij]\{ab\}}+\Psi _{L\{ij\}[ab]},\text{ \ }%
45=(10,\text{ }3)+(15,\text{ }1)  \label{sl2}
\end{equation}%
Here, we have used that any common antisymmetry across two or more combined $%
SL(10,C)$ indices ($ia$,\ $jb$, $kc$) implies antisymmetry in the $SU(5)$
indices ($i,j,k=1,...,5$) and symmetry in the spinor indices ($a,b,c=1,2$),
and the reverse holds as well (the dimensionality of the representations is
also indicated). Notably, while the fermionic $SU(5)$ antiquintet is easily
constructed (\ref{sl1}), the fermionic decuplet does not arise from the
purely antisymmetric $SL(10,C)$ representation (\ref{sl2}). Instead, the
tensor in (\ref{sl2}) corresponds to a collection of vector and scalar
multiplets rather than fermionic ones. This effectively rules out the
standard $SU(5)$ GUT, along with its supersymmetric \cite{su} or
string-inspired \cite{98, 99} extensions from consideration.

Note, in this connection, that all GUTs assigning fermions to purely
antisymmetric representations appear to be irrelevant, as the spin
magnitudes of the resulting states do not match those observed in reality.
The most well-known example of this kind is the $SU(11)$ GUT \cite{ge} with
all three quark-lepton families collected in its one-, two-, three-, and
four-index antisymmetric representations. No doubt, this GUT should also be
excluded in the framework of the considered $SL(2N,C)$ theories. In fact, to
ensure the correct spin-$1/2$ assignment for ordinary quarks and leptons,
these theories must incorporate more intricate fermion multiplets, which
generally feature both upper and lower indices rather than being purely
antisymmetric. However, such multiplets tend to be excessively large and
typically include numerous exotic states that have never been observed.

This may prompt further exploration into the composite structure of quarks
and leptons, for whose constituents --- preons --- the $SL(2N,C)$
unification may appear significantly simpler. As we demonstrate below, the
application of $SL(2N,C)$ symmetry to a model of composite quarks and
leptons --- where chiral preons reside in the fundamental multiplets of $%
SL(2N,C)$, while their massless composites form one of its third-rank
representations --- identifies $SL(16,C)$ HUT as the most likely candidate
for the hyperunification of existing elementary forces, accommodating all
three quark-lepton families \cite{jpl}.

\section{$SL(2N,C)$ unification for composite quarks and leptons}

\subsection{Preons - metaflavors and metacolors}

The key elements of the preon model under consideration are as recently
formulated in \cite{ch}.

At small distances, potentially approaching the Planck scale, there exist $2N
$ elementary massless left-handed and right-handed preons $2N$ elementary
massless left-handed and right-handed preons, described by the independent
Weil spinors $P_{iaL}$ and\ $Q_{iaR}$ (where $i=1,\ldots ,N$ and $a=1,2$),
which share a common local metaflavor symmetry, identified as our
hyperunified $SL(2N,C)$ symmetry\footnote{%
By tradition, we call it the "metaflavor" symmetry, while still referring to
the $SU(N)$ subgroup of $SL(2N,C)$ as the hyperflavor symmetry.},
\begin{equation}
G_{MF}=SL(2N,C)  \label{mf}
\end{equation}%
The preons, both $P_{iaL}$ and $Q_{iaR}$, transform under the fundamental
representation of the $SL(2N,C)$ and their metaflavor theory presumably has
an exact $L$-$R$ symmetry. Actually, the $SL(2N,C)$\ appears at the outset
as some vectorlike symmetry which then breaks down at large distances to
some of its chiral subgroup. All observed quarks and leptons are proposed to
consist of these universal preons some combinations of which are collected
in composite quarks, while others in composite leptons.

The preons also possess a local chiral metacolor symmetry
\begin{equation}
G_{MC}=SO(n)^{L}\times SO(n)^{R}  \label{mc}
\end{equation}%
which contains them in its basic vector representation. In contrast to their
vectorlike metaflavor symmetry, the left-handed and right-handed preon
multiplets are taken to be chiral under the metacolor symmetry. They appear
with different metacolors, $P_{iaL}^{\boldsymbol{\alpha }}$ and \ $Q_{iaR}^{%
\boldsymbol{\alpha }^{\prime }}$, where $\alpha $ and $\alpha ^{\prime }\ $%
are indices of the corresponding metacolor groups $SO(n)^{L}$ and $SO(n)^{R}$%
{\ (}$\alpha ,\alpha ^{\prime }$ ${=1,...,n}$), respectively. These chiral $%
SO(n)^{L,R}$ metacolor forces bind preons into composites --- quarks,
leptons and other states. As a consequence, there are two types of
composites at large distances being composed individually from the
left-handed and right-handed preons, respectively. They presumably have a
similar radius of compositeness, $R_{MC}$ $\sim 1/\Lambda _{MC}$, where $%
\Lambda _{MC}$ corresponds to the scale of the preon confinement for the
asymptotically free (or infrared divergent) metacolor symmetries. Due to the
proposed $L$-$R$ invariance, the metacolor symmetry groups $SO(n)^{L}$ and $%
SO(n)^{R}$ are taken identical with the similar scales for both of sets of
preons. The\ choice of the orthogonal symmetry groups for metacolor has some
advantages over the unitary metacolor symmetry commonly used. Indeed, the
orthogonal metacolor symmetry is generically anomaly-free for its basic
vector representation in which preons are presumably located. Also, the
orthogonal metacolor allows more possible composite representations of the
metaflavor symmetry $SL(2N,C)$ including those which are described by\
tensors with mixed upper and lower metaflavor indices.

Under the local symmetries involved, both the metaflavor $SL(2N,C)$ and
metacolor $SO(n)^{L}\times SO(n)^{R}$, the assignment of the left-handed and
right-handed preons looks as
\begin{equation}
P_{iaL}^{\alpha }[N,(1/2,0);(n,1)]\text{ , \ }Q_{iaR}^{\alpha ^{\prime
}}[N,(0,1/2);(1,n)]  \label{as}
\end{equation}%
In fact, they also possess an accompanying chiral global symmetry\footnote{%
Note that the $SU(N)_{L}\times SU(N)_{R}$ is a chiral symmetry of the
independent lefthanded ($P_{iaL}^{\boldsymbol{\alpha }}$) and righthanded ($%
Q_{iaR}^{\boldsymbol{\alpha }^{\prime }}$) Weil spinors rather than chiral
symmetry related to $L$- and $R$-components of the same Dirac spinors, as is
usually implied.}
\begin{equation}
K(N)=SU(N)_{L}\times SU(N)_{R}  \label{ch}
\end{equation}%
in the limit when their common gauge $SL(2N,C)$ metaflavor interactions are
switched off.

Obviously, the preon condensate $\left\langle \overline{P}%
_{L}Q_{R}\right\rangle $ which could cause the metacolor scale $\Lambda
_{MC} $ order masses for composites is principally impossible in the
left-right metacolor model. This may be generally considered as a necessary
but not yet a sufficient condition for masslessness of composites. The
genuine massless fermion composites are presumably only those which preserve
chiral symmetry of preons (\ref{ch}) at large distances that is controlled
by the corresponding anomaly matching (AM) condition \cite{t}. As we see
below, just a preservation of the chiral symmetry $K(N)$ determines the
particular metaflavor symmetry $SL(2N,C)$, which, in whole or in part, could
be observed at large distances through the massless composites that emerge.

\subsection{Composites -- anomaly matching condition}

An actual metaflavor theory in our model is based on a local vectorlike $%
SL(2N,C)$\ symmetry unifying $N$ left-handed and $N$\ right-handed preons,
while their chiral symmetry $SU(N)_{L}\times SU(N)_{R}$ is in fact global.
However, one can formally turn it into the would-be local symmetry group
with some spectator gauge fields and fermions \cite{t} to properly analyze
the corresponding gauge anomaly cancellation thus checking the chiral
symmetry preservation for massless preons and composites at both small and
large distances. It is important to see that, whereas in the $SL(2N,C)$
metaflavor theory gauge anomalies of preons and composites are automatically
cancelled out between left-handed and right-handed states involved, in the
spectator gauge $SU(N)_{L}\times SU(N)_{R}$ theory all anomalies have to be
cancelled by special multiplets of the metacolorless spectator fermions
introduced individually for the $SU(N)_{L}$ and $SU(N)_{R}$ sectors of the
theory. Though the $SL(2N,C)$ metaflavor interactions may in principle break
the preon chiral symmetry (\ref{ch}), they are typically too weak to
influence the bound state spectrum.

In the proposed preon model the AM condition \cite{t}\ states, in general,
that the chiral $SU(N)_{L}^{3}\ $ and $SU(N)_{R}^{3}$ triangle anomalies
related to $N$ left-handed and $N$ right-handed preons have to match those
for massless composite fermions being produced by the $SO(n)^{L}$ and $%
SO(n)^{R}$ metacolor forces, respectively. Actually, fermions composed from
the left-handed preons and those composed from the right-handed ones have to
independently satisfy their own AM conditions. In contrast, in the local $%
SL(2N,C)$ metaflavor theory being as yet vectorlike, the $SL(2N,C)^{3}$
metaflavor triangle anomalies of the $L$-preons and $R$-preons, as well as
anomalies of their left-handed and right-handed composites, will
automatically compensate each other for any number $N$ of the starting preon
species. However, the AM condition through the constraints on the admissible
chiral symmetry $SU(N)_{L}\times SU(N)_{R}$ providing the masslessness of
composite fermions at large distances, may put in general a powerful
constraint on this number, and thereby on the underlying local metaflavor
symmetry $SL(2N,C)$ itself as a potential local symmetry of massless (or
light) composites. This actually depends on the extent to which the
accompanying global chiral symmetry (\ref{ch}) of preons remains at large
distances.

In one way or another, the AM condition
\begin{equation}
na(N)=\sum_{r}i_{r}a(r)  \label{am}
\end{equation}%
for preons (the left side) and composite fermions (the right side) should be
satisfied. For the sake of brevity, the equation (\ref{am}) is
simultaneously written for both the left-handed and right handed preons and
their composites. Here $a(N)$ and $a(r)$ are the group coefficients of
triangle anomalies related to the groups $SU(N)_{L}$ or $SU(N)_{R}$ in (\ref%
{ch}) whose coefficients are calculated in an ordinary way, \ \ \ \
\begin{equation}
a(r)d^{ABC}=\mathrm{Tr}(\{T^{A}T^{B}\}T^{C})_{r}\text{ }  \label{a}
\end{equation}%
where $T^{A}$ ($A,B,C=1,\ldots ,N^{2}-1$) are the $SU(N)_{L,R}$ generators
taken in the corresponding representation $r$. The $a(N)$ corresponds to a
fundamental representation and is trivially equal to $\pm 1$ (for
left-handed and right-handed preons, respectively), while $a(r)$ is related
to a representation $r$ for massless composite fermions. The value of the
factors $i_{r}$ give a number of times the representation $r$ appears in a
spectrum of composite fermions and is taken positive for the left-handed
states and negative for the right-handed ones. The anomaly coefficients for
composites $a(r)$ contain an explicit dependence on the number of preons $N$%
, due to which one could try to find this number from the AM condition taken
separately for the $L$- and $R$-preons and their composites. In general,
there are too many solutions to the condition (\ref{am}) for any value of $N$%
. Nevertheless, for some special, though natural, requirements an actual
solution may only appear for $N=8$, as we will see below.

\subsection{ $SL(16,C)$ hyperunification and emergent $SL(2,C)\times SU(8)$}

To strengthen the AM condition, one may impose certain constraints on the
spectrum of massless composites expected at large distances. These
constraints, which primarily ensure minimality, are the following three:

($i$) All the massless composites contributing into the AM condition (\ref%
{am}) are spin-$1/2$ states;

($ii$) There are only the minimal three-preon fermion composites which are
formed by the $SO(3)^{L}$ and $SO(3)^{R}$ metacolor forces, respectively;

($iii$) All massless composites emerge within a single representation of the
hyperunified symmetry group $SL(2N,C)$, rather than being distributed across
multiple representations.

We now examine the impact of each constraint on the composite model of
quarks and leptons. Regarding the first constraint ($i$), one may note that
the AM condition (\ref{am}), which involves spin-$1/2$ preons on the left
side, can hardly be maintained across the entire $SL(2N,C)$ symmetry. This
is because any composite multiplet in the theory will generally include, in
addition to spin-$1/2$ states, some higher-spin states as well. The only
natural way for the AM condition to hold is if constraint ($i$) remains in
force. However, this implies that the preon metaflavor symmetry $SL(2N,C)$
must reduce to its $SL(2,C)\times SU(N)$ subgroup at large distances where
the composites emerge. Below, we provide a specific example of this
reduction.

Notably, this symmetry reduction aligns with our earlier findings, albeit
from a different perspective. Specifically, the $SL(2N,C)$ gauge theory,
when formulated with neutral tetrads, effectively reduces to an $%
SL(2,C)\times SU(N)$ invariant framework, where only the vector field
multiplet and a singlet tensor field dynamically persist in the particle
spectrum. In the composite model, this outcome arises independently as a
consequence of the preserved chiral symmetry $SU(N)_{L}\times SU(N)_{R}$ at
large distances, ultimately yielding a theory with the residual metaflavor
symmetry $SL(2,C)\times SU(N)$.

As to the second and third constraints, ($ii$) and ($iii$), all the
three-preon massless composites formed by the metacolor forces emerge within
a single representation of the hyperunified symmetry group $SL(2N,C)$, or,
if the first constraint ($i$) is assumed, within a single representation of
its residual metaflavor symmetry $SL(2,C)\times SU(N)$. While this would not
significantly impact the gauge sector of the theory, it could render its
Yukawa sector --- particularly for composite quarks and leptons ---
considerably less arbitrary. In this way, the strengthened AM condition
suggests that only the single $SL(2,C)\times SU(N)$ multiplets for the
left-handed and right-handed composite, each transforming under the
respective representations of the chiral symmetry groups $SU(N)_{L,R}$ in (%
\ref{ch}), may contribute to the anomaly matching condition (\ref{am}).
Clearly, to ensure the $SL(2N,C)$ metaflavor theory remains anomaly-free,
these $L$-preon and $R$-preon composite multiplets must be similar.

To ensure chiral symmetry preservation, we impose that some\ single $%
SU(N)_{L,R}$ representation $r_{0}$ for massless three-preon states
satisfies the AM condition (\ref{am}). This requirement leads to the
equation
\begin{equation}
3=a(r_{0})  \label{am1}
\end{equation}%
for $n=3$, $a(N)=1$ and $i_{r}=i_{r_{0}}=1$ applied individually for $L$%
-preon and $R$-preon composites. The possible spin-$1/2$ composites formed
from three $L$-preons and three $R$-preons correspond to the third-rank
tensor representations of $SU(N)_{L,R}$
\begin{equation}
\Psi _{\{ijk\}L,R}\text{ },\text{ }\Psi _{\lbrack ijk]L,R}\text{ },\text{ }%
\Psi _{\{[ij]k\}L,R}\text{ },\text{ }\Psi _{\{jk\}L,R}^{i}\text{ },\text{ }%
\Psi _{\lbrack jk]L,R}^{i}  \label{tens}
\end{equation}%
where spinor indices are omitted for brevity.

Now, by calculating their anomaly coefficients and substituting them into
the AM condition (\ref{am1}) one can readily confirm that an integer
solution for $N$ exists only for the tensors $\Psi _{\lbrack jk]L}^{i}$ and $%
\Psi _{\lbrack jk]R}^{\prime i}$. This corresponds to the unique "eightfold"
solution
\begin{equation}
3=N^{2}/2-7N/2-1,\text{ }N=8\text{ .}  \label{8-1}
\end{equation}%
fixing the dimensions of these tensors as $216_{L,R}$. The solution (\ref%
{8-1}) implies that out of all possible chiral symmetries $K(N)$ in (\ref{ch}%
) only the $SU(8)_{L}\times SU(8)_{R}$ symmetry can, in principle, ensure
masslessness of left-handed and\ right-handed fermion composites at large
distances\footnote{%
Generally, due to direct screening effects of the orthogonal metacolor the
massless three-preon composites may appear for any odd number of metacolors
rather than in the $n=3$ metacolor case only. Indeed, one can easily check
that for any higher odd $n$ value the AM condition also works for the
three-preon tensors $\Psi _{\lbrack jk]L}^{i}$ ($\Psi _{\lbrack
jk]R}^{\prime i}$) extended by some number $p$ of one-preon multiplets $\Psi
_{iL}$ ($\Psi _{iR}^{\prime }$)
\begin{equation*}
\Psi _{\lbrack jk]L}^{i}+p\Psi _{iL}\text{ , \ }\Psi _{\lbrack jk]R}^{\prime
i}+p\Psi _{iR}^{\prime }\text{ }
\end{equation*}%
provided that they all are properly screened by the corresponding
metagluons. Actually, the AM condition now leads to the equation
generalizing the anomaly matching condition (\ref{8-1})%
\begin{equation*}
N^{2}/2-7N/2-1+p=n
\end{equation*}%
One can see that there appear solutions only for $n-p=3$ and, therefore, one
has again solution for the eightfold chiral symmetry $SU(8)_{L}\times
SU(8)_{R}$ even for a general $SO(n)^{L}\times SO(n)^{R}$ metacolor case.}.
This, in turn, suggests that among all metaflavor $SL(2N,C)$ symmetries only
$SL(16,C)$ is the most plausible candidate for hyperunification. Note that,
in contrast to the global chiral symmetry described above, in the local $%
SL(16,C)$ metaflavor theory, which as yet remains vectorlike, all metaflavor
triangle anomalies are automatically cancelled.

We finally illustrate how the hyperunified symmetry reduces, when moving
from the spectator theory with chiral $SU(8)_{L}\times SU(8)_{R}$\ symmetry
and its composite multiplets $\Psi _{\lbrack i\text{ }j]L,R}^{k}$ satisfying
the AM condition, to the\ $SL(16,C)$ metaflavor theory. Consider, for
example, one of its multiplets, $\Psi _{\lbrack ia,\text{ }jb]L,R}^{kc}$,
treated as the single representation for the left-handed and right-handed
composite fermions at large distances. In terms of the $SU(8)\times SL(2,C)$
component these multiplets are expressed as follows%
\begin{eqnarray}
\Psi _{\lbrack ia,\text{ }jb]L,R}^{kc} &=&\left( \Psi _{\lbrack
ij]\{ab\}}^{kc}+\Psi _{\{ij\}[ab]}^{kc}\right) _{L,R}\text{\ }  \notag \\
1904_{L,R} &=&[(216,2)+(216+8,\text{ }4)+(280+8,\text{ }2)]_{L,R}
\label{216C}
\end{eqnarray}%
where they are decomposed into symmetric and antisymmetric parts over the
spinor indices, with the dimensions of all its spin-$1/2$ and spin-$3/2$
submultiplets indicated. One can immediately verify that among all
submultiplets in (\ref{216C}) only the $(216,2)_{L,R}$ --- corresponding to
the above tensors $\Psi _{\lbrack jk]L,R}^{i}$ in (\ref{tens}) of the chiral
$SU(8)_{L,R}$ \ symmetries --- individually satisfy the anomaly matching
condition (\ref{8-1}). Consequently, all other submultiplets there must
acquire superheavy masses. This results in a theory with the residual
metaflavor symmetry $SL(2,C)\times SU(8)$ at large distances, where
composite fermions emerge, while the full $SL(16,C)$ symmetry remains intact
for preons at small scales.

\subsection{$L$-$R$ symmetry violation -- composite quarks and leptons}

The survived massless multiplet $\Psi _{\lbrack jk]aL,R}^{i}$ of the
residual metaflavor symmetry $SL(2,C)\times SU(8)$, when decomposed into the
$SU(5)\times SU(3)$ components take the form (spinor indices are omitted)%
\begin{equation}
216_{L,R}=[(\overline{5}+10,\text{ }\overline{3}%
)+(45,1)+(5,8+1)+(24,3)+(1,3)+(1,\overline{6})]_{L,R}  \label{216}
\end{equation}%
where the first term in the squared brackets, for left-handed states in $%
216_{L}$, describes all three quark-lepton families as the corresponding
triplets of the $SU(3)$ family symmetry \cite{h}. However, the same
structure also appears for right-handed states in $216_{R}$, reflecting the
vectorlike nature of our $SL(16,C)$ theory. This implies that while preons
remain massless, protected by their own metacolor symmetries, the
metacolor-singlet composites in (\ref{216}) can pair up and acquire heavy
Dirac masses.

To prevent such an occurrence for the physical quark and lepton submultiplet
in (\ref{216}), $(\overline{5}+10,$ $\overline{3})_{L}$, one may follow the
recent scenario \cite{ch} introducing a spontaneous breaking of the basic $L$%
-$R$ symmetry in the theory. This breaking is assumed to originate from the
interactions of right-handed preons, which reduces the chiral symmetry of
their composites down to $[SU(5)\times SU(3)]_{R}$ at large distances. In
fact, such breaking may naturally arise due to the possible $L$-$R$
asymmetric condensation of massive composite scalars, which inevitably
appear alongside composite fermions. As a result, although massless
right-handed preons still respect the full $SU(8)_{R}$ symmetry, the
masslessness of their composites is now governed solely by its remaining $%
[SU(5)\times SU(3)]_{R}$ subgroup. Consequently, while the left-handed preon
composites continue to complete the entire $216_{L}$ in (\ref{216}), the
right-handed preon composites, constrained by their residual chiral
symmetry, no longer include all the submultiplets present in $216_{R}$.
Remarkably, the corresponding anomaly matching condition organizes the
composite spectrum in such a way that the submultiplet $(\overline{5}+10,$ $%
\overline{3})_{R}$ is absent among the right-handed preon composites. As a
result, all the left-handed submultiplet members in $(216,2)_{L}$, except
the $(\overline{5}+10,$ $\overline{3})_{L}$, pair up, becoming heavy and
decoupling from laboratory physics \cite{ch}.

Accordingly, once the $L$-$R$ symmetry is violated in the theory, the
vectorlike metaflavor symmetry $SU(8)\times SL(2,C)$ breaks down to its
chiral subgroup $[SU(5)\times SU(3)_{F}]\times SL(2,C)$ for the
large-distance composites. This ultimately leads to the conventional $SU(5)$
GUT combined with the chiral $SU(3)_{F}$ family symmetry \cite{h},
describing the three standard families of composite quarks and leptons.

\subsection{Down to Standard Model}

The further symmetry violation is related, as was mentioned above, to the
adjoint scalar field multiplet $\Phi $ (\ref{ad}) which in the present
context breaks the $SU(5)$ to the Standard Model.

As to the final breaking of the SM and accompanying family symmetry $%
SU(3)_{F}$, it occurs through the extra multiplets $H^{[ia,jb,kc,ld]}$, and $%
\chi _{\lbrack ia,jb]}$ and $\chi _{\{ia,jb\}}$ of $SL(16,C)$, which contain
an even number of antisymmetrized and symmetrized indices. These multiplets
include, among others, the genuine scalar components which develop the
corresponding VEVs giving masses to the weak bosons, as well as to the
family bosons of the $SU(3)_{F}$. They also generate masses for quarks and
leptons located in the left-handed fermion multiplet (\ref{216}) through the
$SL(16,C)$ invariant Yukawa couplings%
\begin{eqnarray}
&&\frac{1}{\emph{M}}\left[ \Psi _{\lbrack ja,\text{ }kb]L}^{ic}C\Psi
_{\lbrack me,\text{ }nf]L}^{ld}\right] H^{\{[ja,kb],[me,nf]\}}(a_{u}\chi
_{\lbrack ic,ld]}+b_{u}\chi _{\{ic,ld\}})  \notag \\
&&\frac{1}{\emph{M}}\left[ \Psi _{\lbrack ja,\text{ }kb]L}^{ic}C\Psi
_{\lbrack ic,\text{ }me]L}^{ld}\right] H^{\{[ja,kb],[me,nf]\}}(a_{d}\chi
_{\lbrack ld,nf]}+b_{d}\chi _{\{ld,nf\}})  \label{yc}
\end{eqnarray}%
with distinct index contraction for the up quarks, and down quarks and
leptons, respectively ($i,j,k,l,m,n=1,...,8;$ $a,b,c,d,e,f=1,2$). The mass $%
\emph{M}$ represents an effective scale in the theory that, in the composite
model of quarks and leptons, can be linked to their compositeness scale,
while $a_{u,d}$ and $b_{u,d}$ are dimensionless constants of the order of $1$%
.

Actually, these couplings contain two types of scalar multiplets with the
following $SU(8)\times SL(2,C)$ components -- the $H$ multiplet $%
H^{\{[ja,kb],[me,nf]\}}$ containing the true scalar components%
\begin{equation}
H^{[jkmn]\{[ab],[ef]\}}(70,1)  \label{70}
\end{equation}%
and the symmetric and antisymmetric $\chi $ multiplets, $\chi _{\{ic,ld\}}$
and $\chi _{\lbrack ic,ld]}$, whose scalar components look as%
\begin{equation}
\chi _{\lbrack il\boldsymbol{]}[cd]}(28,1),\text{ \ }\chi _{\lbrack cd]\{il%
\boldsymbol{\}}}(36,1)  \label{28}
\end{equation}%
Decomposing them into the components of the final $SU(5)\times SU(3)_{F}$
symmetry one finds the full set of scalars
\begin{eqnarray}
70 &=&(5,1)+(\overline{5},1)+(10,\overline{3})+(\overline{10},3)  \notag \\
28 &=&(5,3)+(10,1)+(1,\overline{3})  \notag \\
36 &=&(5,3)+(15,1)+(1,6)  \label{36}
\end{eqnarray}%
containing the $SU(5)$ quintets $(5,1)$ and $(\overline{5},1)$ to break the
Standard Model at the electroweak scale $M_{SM}$, and the $SU(3)_{F}$
triplet and sextet, $(1,\overline{3})$ and $(1,6)$, to properly break the
family symmetry at some larger scale $M_{F}$. One may refer to the scalars (%
\ref{70}) and (\ref{28}) as the respective "vertical" and "horizontal"
scalars, which provide the simplest form of the above Yukawa couplings.
Acting in pairs, they presumably determine the masses and mixings of all
quarks and leptons. Lastly, and importantly, in the model under
consideration, these scalars may themselves be composite states formed from
the same preons as quarks and leptons \cite{ch}.

We may conclude that the composite structure of quarks and leptons
introduces a new fundamental scale in the theory, in addition to the Planck
scale $M_{P}$ and the $SL(2N,C)$ breaking scale $M$ ---the preon confinement
scale $\Lambda _{MC}$ inside quarks and leptons. All consequent low-energy
physics will depend on the interplay of these three scales. It is possible
that they nearly coincide; in that case, predictions would largely collapse
to an a posteriori explanation of present observations, adding nothing new.
By contrast, if $M$ and $\Lambda _{MC}$ are significantly lower than $M_{P}$%
, many new phenomena may emerge, potentially yielding direct or indirect
confirmation of the relevance of $SL(2N,C)$ hyperunification in nature.

\section{Summary and outlook}

We have investigated the potential of the local $SL(2N,C)$ symmetry to unify
all fundamental forces, including gravity. The view that gravity alone
dictates the symmetry structure required for this hyperunification naturally
leads from the $SL(2,C)$ gauge\ gravity to the broader local symmetry $%
SL(2N,C)$ with $N$ defining the internal $SU(N)$ symmetry subgroup. In this
setting, aside from the linear gravity Lagrangian---to which only tensor
fields contribute---the quadratic sector is unified across all gauge
submultiplets with a single universal gauge coupling.

Tetrads play a central role throughout $SL(2N,C)$. Once made dynamical,
their orthonormality condition naturally implies condensation, which can be
viewed as a nonlinear sigma--model--type length constraint. The effective
symmetry thus reduces to $SL(2,C)\times SU(N)$, collecting together $SL(2,C)$
gauge gravity and an $SU(N)$ grand-unified sector, and thereby evading the
assumptions behind the Coleman--Mandula theorem.

The algebraic enlargement from $SL(2,C)$ to $SL(2N,C)$ introduces no extra
spacetime dimensions; rather, it unifies the spinor structure and internal
symmetries within a single gauge framework. Consequently, while the full
gauge multiplet of $SL(2N,C)$ encompasses vector, axial-vector and tensor
field submultiplets, only the vector field submultiplet emerge in the
observed particle spectrum. The axial-vector and tensor field submultiplets
acquire heavy masses at the $SL(2N,C)$ symmetry breaking scale triggered by
the tetrad condensation.

On the gravitational side, the extension of $SL(2,C)$ gauge gravity to $%
SL(2N,C)$ hyperunification calls for \textquotedblleft
safe\textquotedblright\ curvature-squared terms alongside the standard
quadratic terms for gauge strengths. This criterion uniquely identifies,
from among all possible candidates, the Neville ghost-free curvature-squared
Lagrangian \cite{nev, nev1}, as the most appropriate model for such an
extension. Consequently, the resulting theory includes the $SL(2N,C)$
symmetric $R+R%
{{}^2}%
$ Einstein-Cartan type gravity action, which remains free from ghosts and
tachyons. The theory also contains the properly suppressed four-fermion
(spin current-current) interaction (\ref{6}) which, in contrast to the
standard case \cite{Kibble}, includes the hyperflavor depending interaction
terms as well.

For the grand unification, in turn, since all states involved in $SL(2N,C)$
theories are additionally classified by spin magnitude, the $SU(N)$ GUTs
with purely antisymmetric matter multiplets, including the usual $SU(5)$
theory, turn to be irrelevant for the standard $1/2$-spin quarks and
leptons. Meanwhile, the $SU(8)$ grand unification, arising from the $SL(16,C)
$ theory, originally introduced for preon constituents, successfully
describes all three families of composite quarks and leptons. This leads to
a theory with the residual metaflavor symmetry $SL(2,C)\times SU(8)$ for
composite fermions, while the full $SL(16,C)$ symmetry remains intact for
preons at small scales. In general, $SL(2N,C)$ appears to call for a deeper
level of elementarity in the existing particle spectrum at which the
framework can operate successfully.

Axial-vector fields---even if massive--- may still pose a challenge within
the theory. However, in the neutral-tetrad setting emerged, axial-vector
field multiplet, unlike vector and tensor fields, do not directly couple to
fermionic matter. This may point to a novel scenario for hyperunification,
where all gauge fields associated with the $SL(2N,C)$ symmetry arise as
composite bosons built from fermion bilinears rather than elementary fields.
This approach, long recognized as a viable alternative to conventional
quantum electrodynamics \cite{bjorken}, gravity \cite{ph} and Yang-Mills
theories \cite{eg, ter, suz, jpl2}, has not yet been explored in the context
of noncompact unified symmetries. In such a framework, where only the global
$SL(2N,C)$ symmetry imposed on a purely fermionic (preonic) Lagrangian and
suitably constrained currents, one expects composite vector and tensor
fields to emerge dynamically, while axial-vector modes do not.

Further study may also focus on the phenomenological aspects of the theory.
The breaking of the $SL(2N,C)$ HUT to the effective $SL(2,C)\times SU(N)$
symmetry and further down to the Standard Model, gives rise to numerous new
processes. These emerge from the generalized gravity and Standard Model
sectors, leading to novel particles and interactions. We also anticipate a
characteristic series of massive fermionic and bosonic composites produced
by preon dynamics at the confinement scale, provided this scale becomes
observationally accessible in the near future.

These important questions will be addressed elsewhere.

\section*{Acknowledgments}

I am grateful to Colin Froggatt, Oleg Kancheli and Holger Nielsen for
illuminating discussions.

%
%
%
%
%
%
%
%
%
\thispagestyle{empty}

\end{document}